\documentclass[aps,prc,twocolumn,superscriptaddress,floatfix,showpacs,nofootinbib]{revtex4-1}
\usepackage{multirow}
\usepackage{graphicx}
\usepackage{color}
\usepackage{amsmath}
\usepackage{amssymb}
\usepackage[colorlinks=true]{hyperref}
\usepackage{tikz}
\usepackage{etoolbox}
\usepackage{bm}
\usepackage[utf8]{inputenc}
\usepackage{cleveref}
\usepackage{amsmath}
\usepackage{amssymb}
\usepackage{verbatim}
\usepackage{float}
\usepackage[bottom]{footmisc}
\usepackage[justification=raggedright]{caption}
\usepackage{subcaption}
\usepackage{verbatim}
\usepackage{comment}
\usepackage{booktabs}
\usepackage{array}
\usepackage{combelow}
\usepackage{ulem}
\newcommand{\SVH}    {{scaled {\tt vHLLE}}}
\newcommand {\RTA}   {{{\tt RTA} kinetic theory}}
\newcommand {\idl} {\mathrm{ideal}}
\definecolor{dgreen}{cmyk}{1.,0.,1.,0.4}        % dark green
\definecolor{orange}{cmyk}{0.,0.353,1.,0.}    % orange

\newcommand{\mr}[1]{\mathrm{#1}}
\newcommand{\mc}[1]{\mathcal{#1}}

\newcommand{\lrp}[1]{\left(#1\right)}
\newcommand{\lra}[1]{\left\langle{#1}\right\rangle}

\newcommand{\xT}{\mathbf{x}_\perp}

\newcommand {\VAH}   {{\tt VAH}}
\newcommand {\RTa}   {{\tt RTA}}
\newcommand {\vHLLE}   {{\tt vHLLE}}

\begin{document}
%\begin{CJK*}{GB}{}
\title{
%Can Anisotropic Hydrodynamics Expand the Domain of Applicability of Hydrodynamics in 2+1D evolution? {\color{blue}(Anisotropic Hydrodynamics Expands the Domain of Applicability of Hydrodynamics in Systems with Transverse Dynamics)}
%{\color{purple} [VEA] 
Extended applicability domain of viscous anisotropic hydrodynamics in $(2{+}1)$-D Bjorken flow with transverse expansion
%}
}

\author{Yiyang Peng}
\email{yiyangpeng@stu.pku.edu.cn}
\affiliation{School of Physics, Peking University, Beijing 100871, China}

\author{Victor E. Ambru\cb{s}}
\email{victor.ambrus@e-uvt.ro}
\affiliation{Department of Physics, West University of Timi\cb{s}oara, Bd. Vasile P\^arvan 4, Timi\cb{s}oara 300223, Romania}

\author{Clemens Werthmann}
\email{clemens.werthmann@ugent.be}
\affiliation{Department of Physics and Astronomy, Ghent University, 9000 Ghent, Belgium}

\author{S\"{o}ren Schlichting}
\email{schlichting@physik.uni-bielefeld.de}
\affiliation{Fakult\"{a}t f\"{u}r Physik, Universit\"{a}t Bielefeld, D-33615 Bielefeld, Germany}

\author{Ulrich Heinz}
\email{heinz.9@osu.edu}
\affiliation{Department of Physics, The Ohio State University, Columbus, OH 43210-1117, USA}

\author{Huichao Song}
\email{huichaosong@pku.edu.cn}
\affiliation{School of Physics, Peking University, Beijing 100871, China}
\affiliation{Center for High Energy Physics, Peking University, Beijing 100871, China}
\date{\today}
\begin{abstract}
We perform (2{+}1)-D simulations of viscous anisotropic hydrodynamics (\VAH) under boost-invariant and conformal conditions. Comparing both \VAH\ and traditional viscous hydrodynamics with kinetic theory in the relaxation-time approximation as the underlying microscopic theory, we show that \VAH\ provides a superior description of the evolution across a wide range of opacity, effectively extending the applicability of hydrodynamic modeling. Our results demonstrate \VAH's potential for describing collective flow in small systems where traditional hydrodynamics faces challenges.
\end{abstract}


%\pacs{25.75.Ld, 25.75.Gz}

\maketitle
%\end{CJK*}

%-==========================================================================
\clearpage

\section{Introduction}
Relativistic hydrodynamics is an important tool to describe the evolution of the quark-gluon plasma (QGP) and interpret the collective flow in relativistic heavy-ion collisions~\cite{Kolb:2003dz,Teaney:2009qa,Heinz:2013th,Hirano:2012kj,Florkowski:2017olj,Song:2017wtw,Heinz:2024jwu}. The validity of hydrodynamics relies on a clear separation between the macroscopic and microscopic scales, as well as sufficient time to achieve local equilibrium. While this requirement is typically satisfied in sufficiently high multiplicity events, hydrodynamics might be challenged in smaller, more dilute, and short-lived systems. Over the past decade, collective behavior has been observed consistently in small collision systems, such as p+p and p+Pb collisons at the Large Hadron Collider (LHC) and $\text{p}/\text{d}/{}^3\text{He}$+Au at the Relativistic Heavy Ion Collider (RHIC)~\cite{PHENIX:2017xrm,PHENIX:2018lia,ALICE:2014dwt,ATLAS:2017hap,ALICE:2019zfl,CMS:2016fnw,CMS:2017kcs}. Hydrodynamic simulations provide a qualitative but imperfect description of the flow data measured in experiment~\cite{Mantysaari:2025tcg,Shen:2016zpp,Bozek:2011if,Bzdak:2013zma,Qin:2013bha,Nagle:2013lja,Werner:2013tya,Werner:2013ipa,Bozek:2013ska,Schenke:2014zha,
Bozek:2014cya,Bozek:2015swa,Zhou:2015iba,Weller:2017tsr,Mantysaari:2017cni,
Zhao:2017rgg,Schenke:2020mbo,Schenke:2019pmk,OrjuelaKoop:2015etn,Bozek:2015qpa, Zhao:2020pty,Zhao:2020wcd,Zhao:2022ayk}. 
Therefore, understanding the emergence of hydrodynamization and the applicability range of hydrodynamic descriptions is crucial for evaluating flow-like behavior in small collision systems. 

Significant efforts have been devoted to addressing this issue in kinetic theory~\cite{Denicol:2014tha,Kurkela:2018wud,Kurkela:2019cgr,Kurkela:2019kip,Kurkela:2020wwb,Denicol:2014tha,Denicol:2019lio,Du:2020zqg,Du:2022bel,Blaizot:2021cdv,Ambrus:2022koq,Arslandok:2023utm}. Detailed comparisons of the $(2{+}1)$-D evolution in kinetic theory and hydrodynamics reveal that in dilute systems, transverse expansion has been developed before hydrodynamization~\cite{Ambrus:2022koq,Ambrus:2022qya,Werthmann:2023dvl,Ambrus:2023oyk,Ambrus:2024eqa,Ambrus:2024hks}. In such cases, non-hydrodynamic excitations during the pre-equilibrium evolution have non-negligible effects on final-state observables. 

Alternative models to traditional hydrodynamics are required to understand the bulk evolution and anisotropic flow development of small systems. Among these, anisotropic hydrodynamics (aHydro) is particularly promising \cite{Martinez:2010sc,Florkowski:2012pf,Florkowski:2014txa,Strickland:2014pga,Kasmaei:2019ofu,Behtash:2017wqg,Strickland:2024moq}. The leading order (LO) aHydro is derived from kinetic theory using an anisotropic ansatz of the microscopic distribution function~\cite{Martinez:2012tu, Tinti:2014yya, Alqahtani:2016rth}, to capture the early-time anisotropy in momentum space. To go beyond the leading order, viscous anisotropic hydrodynamics includes the non-cylindrical viscous correction in a linearized manner akin to second-order viscous hydrodynamics~\cite{Bazow:2013ifa,Molnar:2016vvu,McNelis:2018jho}. Through rigorous comparison, aHydro has been demonstrated to be accurately consistent with kinetic theory in $(0{+}1)$-D Bjorken flow evolution \cite{Florkowski:2010cf, Florkowski:2015cba, Bazow:2015cha, Tinti:2015xwa}. Furthermore, a program called \VAH\ for complete (3+1)-D simulation of viscous anisotropic hydrodynamics has been developed and implemented~\cite{McNelis:2021zji}. With properly tuned parameters, \VAH\ successfully reproduces transverse momentum spectra and flow harmonics in Pb+Pb collisions at $2.76\,\mr{TeV}$~\cite{Liyanage:2023nds}, 
provides a good description of $v_{2}\{2\}(p_T)$, $v_{3}\{2\}(p_T)$ and $v_{4}\{2\}(p_T)$ over a wide range of multiplicities and transverse momenta \cite{Heinz:2023kzr}, and for the first time correctly reproduces the experimentally observed negative $c_{2}\{4\}$  in p+p collisions at $13\,\mr{TeV}$~\cite{Zhao:2025jwf}.  

Despite the remarkable progress, comparisons between anisotropic hydrodynamics and kinetic theory were limited to the Bjorken flow or other analytic fluid velocity profiles~\cite{Strickland:2015utc,Chen:2024grb}. It will be insightful to investigate this topic in more general scenarios. Motivated by this gap, we present a systematic comparative study of viscous anisotropic hydrodynamics (\VAH) and kinetic theory in $(2{+}1)$-D evolution models with longitudinal boost invariance, with particular focus on how effectively \VAH\ captures the characteristic transverse dynamics of the system, which is sensitive to the degree of hydrodynamization. This research will shed more light on whether \VAH\ can expand the applicability domain of traditional viscous hydrodynamics.

This paper is organized as follows: In Sec. \ref{sec:model,obs}, we briefly describe the models employed in this work and the observables of interest. In Sec. \ref{sec:evolve}, we compare the proper-time evolution of observables between \VAH\ and kinetic theory, assessing the accuracy of \VAH\ in describing the spacetime evolution. In Sec. \ref{sec:final-state}, we analyze the opacity dependence of final-state observables across \VAH, kinetic theory and traditional viscous hydrodynamics, to demonstrate the capabilities of \VAH\ in modeling small systems. Section~\ref{sec:conc} concludes this paper. We furthermore provide two appendices: Appendix~\ref{appendix A} discusses the details of the Bjorken attractor in the three models considered in this paper, while Appendix~\ref{appendix B} addresses the dissipative effects on the momentum anisotropy, $\varepsilon_p$.

\section{Models and Observables}\label{sec:model,obs}
\subsection{Models}\label{sec:models}

%With $f(x,p)$ taken the isotropic or anisotropic form with the non-equilibrium/ anisotropic contribution, one could obtained the specific form of  $T^{\mu\nu}$ with the ideal and  dissipative part as obtained from the  isotropic or anisotropic tensor decomposition with a chosen frame. 

%The results for kinetic theory and traditional hydrodynamics presented in this paper are identical to those from previous works~\cite{Ambrus:2022koq,Ambrus:2022qya}, and the results for vaHydro are obtained by the VAH code. One can refer for more details on the setup and the simulation codes to Refs.~\cite{Ambrus:2021fej,Ambrus:2022koq,McNelis:2021zji,Karpenko:2013wva}.

In this section, we start from the kinetic theory that describes the evolution of the phase space distribution $f(x,p)$  with the relativistic Boltzmann equation.  We assume that the system consists of massless quasi-particles, associated with the conformal symmetry.  In 
%hydrodynamics,
kinetic theory, the energy-momentum tensor $T^{\mu\nu}$ can be obtained from $f(x,p)$ by $T^{\mu\nu}=\nu_{\mr{eff}} \int \frac{d^3p}{(2\pi)^3E_p}p^\mu p^\nu f(x,p)$, with $\nu_{\mr{eff}}$ representing the number of effective bosonic degrees of freedom. The energy-momentum conservation of $T^{\mu\nu}$, as well as the evolution equations for the dissipative terms in standard and anisotropic viscous hydrodynamics, can be derived from kinetic theory with isotropic or anisotropic phase space distribution. For details, please refer to Ref.~\cite{McNelis:2018jho} and the brief summary below.

%In order to reduce the dimension for the numerical simulations, we assume longitudinal boost-invariance. In this simplified scenario, the system exhibits conformal symmetry, and the phase space distribution depends on longitudinal coordinates only via the difference of the pseudorapidity $y=\mathrm{artanh}(p^z/p^t)$ and the spacetime rapidity $\eta=\mathrm{artanh}(z/t)$. We will thus use Milne coordinates $(\tau,x,y,\eta)$, where $\tau=\sqrt{t^2-z^2}$ and the metric tensor is $g_{\mu\nu}=\mathrm{diag}(1,-1,-1,-\tau^2)$.

\textbf{Kinetic theory with \RTa:} In this paper,  we employ the Boltzmann equation with the relaxation time approximation (\RTa) \cite{Anderson:1974nyl} to describe the evolution of the distribution $f(x,p)$ for massless particles:
\begin{equation}
    \label{eq:RTA}
    p\cdot \partial f(x,p) = -\frac{u\cdot p}{\tau_R(x)}\left[f(x,p)-f_{\mathrm{eq}}(x,p)\right],
\end{equation}
where $\tau_R$ is the relaxation time, related to the shear viscosity $\eta/s$ via $\tau_R=5\eta/(sT)$ \cite{Florkowski:2013lya}. For the remainder of this paper, we consider $\eta/s$ to be constant. In the equilibrium distribution function  $f_{\mathrm{eq}}=1/(e^{p\cdot u / T}{-}1)$, the local temperature $T(x)$ and the flow velocity $u^\mu(x)$ are determined by the Landau matching condition $u_\mu T^{\mu\nu}=\epsilon u^\nu$. The energy density $\epsilon$ satisfies the conformal equation of state (EoS) $\epsilon = aT^4$, where $a=\nu_{\mathrm{eff}}\pi^2/30$ with $\nu_{\mathrm{eff}}=42.25$, compatible with the high-temperature lattice QCD results reported in Refs.~\cite{HotQCD:2014kol,Borsanyi:2016ksw}.

\textbf{Traditional viscous hydrodynamics:} The underlying microscopic distribution function is assumed to be near equilibrium, which can be written into  $f(x,p)= f_{eq}+ \delta f$. Correspondingly, the energy-momentum tensor $T^{\mu\nu}$ in the Landau frame can be decomposed as follows:
\begin{align}\label{eq:tmunu}
T^{\mu\nu}=\epsilon u^\mu u^\nu-(P+\Pi) \Delta^{\mu\nu}+\pi^{\mu\nu}.
\end{align}
where $\Delta^{\mu\nu}=g^{\mu\nu}-u^\mu u^\nu$, $\epsilon$ is the energy density, $P$ is the thermal equilibrium pressure.  $\pi^{\mu\nu}$ and $\Pi$ are the shear stress tensor and the bulk viscous pressure, respectively, associated with the off-equilibrium distribution $\delta f$. For a massless quasi-particle system that satisfies the conformal symmetry, the bulk viscous pressure $\Pi$ vanishes. The thermal pressure $P$ and the energy density $\epsilon$ satisfy $P=\epsilon/3$.

Energy-momentum conservation requires $\partial_\mu T^{\mu\nu}=0$. The evolution equations for the shear stress tensor $\pi^{\mu\nu}$ can be derived from kinetic theory~\cite{Grad:1949zza, Baier:2006um, Baier:2007ix, Betz:2008me, Denicol:2012cn, Denicol:2012es} or constrained from the second law of thermodynamics~\cite{Israel:1979wp, Muronga:2004sf}. They take the form of relaxation equations of second-order Mueller-Israel-Steward type,
\begin{equation}\label{eq:dpi}
    \begin{split}
        \tau_\pi \dot{\pi}^{\langle \mu\nu \rangle} + \pi^{\mu\nu} &= 2\eta \sigma^{\mu\nu} +  
2\tau_\pi \pi^{\langle\mu}_\lambda \omega^{\nu\rangle \lambda} - \delta_{\pi\pi} \pi^{\mu\nu} \theta\\
&\quad  -  
\tau_{\pi\pi} \pi^{\lambda\langle \mu}\sigma^{\nu\rangle}_\lambda + \phi_7 \pi_\alpha^{\langle\mu}\pi^{\nu\rangle\alpha},
    \end{split}
\end{equation}
where  $\dot{A} = u \cdot \partial A$ denotes the comoving derivative, $A^{\langle\mu\nu\rangle} = \Delta^{\mu\nu}_{\alpha\beta} A^{\alpha\beta}$ is the angular brackets notation with $\Delta^{\mu\nu}_{\alpha\beta} = \frac{1}{2} (\Delta^\mu_\alpha \Delta^\nu_\beta + \Delta^\nu_\alpha \Delta^\mu_\beta) - \frac{1}{3} \Delta^{\mu\nu} \Delta_{\alpha\beta}$ ensuring tracelesness, symmetrization and orthogonality to the fluid four-velocity $u^\mu$,
$\sigma^{\mu\nu}=\nabla^{\langle\mu} u^{\nu\rangle}$ is the shear stress tensor, $\omega^{\mu\nu}=\frac{1}{2}(\nabla^\mu u^\nu-\nabla^\nu u^\mu)$ is the vorticity tensor, and $\nabla^\mu = \Delta^{\mu\nu} \partial_\nu$. 
The transport coefficients in Eq.~\eqref{eq:dpi} can be obtained from kinetic theory with \RTa~\cite{Jaiswal:2013npa,Molnar:2013lta,Ambrus:2022vif}:
\begin{align}
  \eta &= \frac{4}{5} \tau_\pi P, &
 \delta_{\pi\pi} &= \frac{4\tau_\pi}{3}, &
 \tau_{\pi\pi} &= \frac{10\tau_\pi}{7}, &  \phi_7 = 0, %\nonumber\\
 \label{eq:hydro_tcoeffs}
\end{align}
and the relaxation time $\tau_\pi$ is identical to the relaxation time in Eq.~\eqref{eq:RTA}, $\tau_\pi=\tau_R$,
In this paper, the traditional viscous hydrodynamics simulations are performed with
the numerical code \vHLLE; please refer to Ref.~\cite{Karpenko:2013wva} for details.

%%%%%%%%%%%%%%%%%%%%%%%%%%%%%%%%%%%%%%%%%%
\textbf{Viscous anisotropic hydrodynamics (\VAH):} 
%%%%%%%%%%%%%%%%%%%%%%%%%%%%%%%%%%%%%%%%%%%
We solve the equations of motion of viscous anisotropic hydrodynamics with the \VAH\ code. In this framework, $T^{\mu\nu}$ is decomposed in the basis $u^\mu$ and $z^\mu$, where $u^\mu$ is the flow velocity in the Landau frame and $z^\mu$ denotes the normalized vector in the longitudinal direction. This decomposition is expressed as 
\begin{align}\label{eq:vahtmn}
T^{\mu\nu}=\epsilon u^\mu u^\nu+P_Lz^\mu z^\nu-P_\perp \Xi^{\mu\nu}+2W_{\perp z}^{(\mu}z^{\nu)}+\pi_\perp^{\mu\nu},
\end{align}
where $\Xi^{\mu\nu}=\Delta^{\mu\nu}+z^\mu z^\nu$ is the transverse projector, $P_L\equiv z_\mu z_\nu T^{\mu\nu}$ and $P_\perp\equiv-\frac12\Xi_{\mu\nu}T^{\mu\nu}$ represent the longitudinal and transverse pressures, respectively. The transverse shear stress tensor $\pi^{\mu\nu}_\perp$ can be extracted by $\pi_\perp^{\mu\nu}\equiv\Xi^{\mu\nu}_{\alpha\beta}T^{\alpha\beta}$ with $\Xi^{\mu\nu}_{\alpha\beta}=\frac12\left(\Xi^\mu_\alpha\Xi^\nu_\beta+\Xi^\mu_\beta\Xi^\nu_\alpha-\Xi^{\mu\nu}\Xi_{\alpha\beta}\right)$, and $W_{\perp z}^\mu\equiv-\Xi^\mu_\alpha T^{\alpha\nu}z_{\nu}$ is the longitudinal momentum diffusion current. The decompositions (\ref{eq:tmunu}) and (\ref{eq:vahtmn}) are related by
\begin{subequations}
    \begin{equation}\label{eq:pidcp}
    \pi^{\mu\nu}=\pi^{\mu\nu}_{\perp} + 2\,W^{(\mu}_{\perp z} z^{\nu)}
  +\frac{1}{3} \big(P_L - P_\perp\big) 
  \big(2z^\mu z^\nu - \Xi^{\mu\nu}\big),
    \end{equation}
    \begin{equation}
            \Pi = \frac{2\, P_\perp + P_L}{3} - P.
    \end{equation}
\end{subequations}

The evolution equations for $P_L$, $P_\perp$, $\pi_\perp^{\mu\nu}$, and $W_{\perp z}^\mu$ are derived from kinetic theory under the assumption that the anisotropic phase-space distribution can be expanded around the Romatschke-Strickland distribution,
\begin{equation}
f_a(x,p)=f_{\mr{eq}}\lrp{\frac{\sqrt{\Omega_{\mu\nu} p^\mu p^\nu}}{\Lambda}},
\end{equation}
with the quadratic form $\Omega_{\mu\nu}p^\mu p^\nu$ parameterized by
\begin{equation}
    \Omega_{\mu\nu}p^\mu p^\nu = m^2 - \frac{\Xi_{\mu\nu}p^\mu p^\nu}{\alpha_\perp^2}+\frac{(z\cdot p)^2}{\alpha_L^2}.
\end{equation}
The parameters $\Lambda,\;\alpha_L$ and $\alpha_\perp$ are determined by matching $\epsilon,\;P_L$ and $P_\perp$. For massless quasi-particles with $m=0$, the parameter $\alpha_\perp$ becomes redundant because $P_\perp$ can be inferred from the condition $P_L+2P_\perp=\epsilon$ and is thereby fixed to 1. With these simplifications, the small correction $\delta \tilde f=f-f_a$ generates dissipative flows that can be written as a linear combination of $\pi_{\perp}^{\mu\nu}$ and $W_{\perp z}^\mu$~\cite{McNelis:2018jho}. In \VAH\ simulations with a lattice QCD EoS, an evolving mean field $B$ is introduced to maintain thermodynamic consistency \cite{Alqahtani:2016rth}. However, in the massless scenario, this mean field and the bulk dissipative terms in the equations of $P_{L}$ and $P_{\perp}$  are canceled. Furthermore, the longitudinal momentum diffusion current $W_{\perp z}^\mu$ vanishes in the (2+1)-D case with longitudinal boost invariance. Consequently, the equations of \VAH\ for the dissipative quantities are simplified as  
\begin{subequations}
    \begin{equation}\label{eq:VAH PL}
        \dot P_L = -\frac{P_L - P_\perp}{3\tau_\pi/2}+\bar{\zeta}^L_z\theta_L +\bar{\zeta}_\perp^L\theta_\perp -\bar{\lambda}_\pi^L \pi_\perp^{\mu\nu}\sigma_{\perp,\mu\nu},
    \end{equation}
    \begin{equation}
        \dot P_\perp = \frac{P_L-P_\perp}{3\tau_\pi}+\bar{\zeta}_z^\perp\theta_L+\bar{\zeta}_\perp^\perp\theta_\perp+\bar\lambda_\pi^\perp\pi_\perp^{\mu\nu}\sigma_{\perp,\mu\nu},
    \end{equation}
    \begin{equation}\label{eq:VAH piperp}
        \begin{split}
            \dot{\pi}_\perp^{\{\mu\nu\}} =& -\frac{\pi^{\mu\nu}_\perp}{\tau_\pi} +2\bar{\eta}_\perp\sigma^{\mu\nu}_\perp - \bar{\lambda}^\pi_\pi\pi^{\mu\nu}_\perp\theta_L -\bar{\delta}_\pi^\pi\pi_\perp^{\mu\nu}\theta_\perp \\
            &-\bar{\tau}^\pi_\pi\pi_\perp^{\lambda\{\mu}\sigma^{\nu\}}_{\perp,\lambda}+2\pi^{\lambda\{\mu}\omega^{\nu\}}_{\perp,\lambda},
        \end{split}
    \end{equation}
\end{subequations}
where 
%$\dot{P}_L=u\cdot \partial P_L$, 
$\theta_L=-z_\mu z_\nu \partial^\mu u^\nu$, $\theta_\perp=\Xi_{\mu\nu}\partial^\mu u^\nu$, $\sigma_\perp^{\mu\nu}=\Xi^{\mu\nu}_{\alpha\beta}\partial^\alpha u^\beta$, $\omega_{\perp}^{\mu\nu}=\frac12\Xi^\mu_\alpha\Xi^\nu_\beta(\partial^\alpha u^\beta - \partial^\beta u^\alpha)$ and the curly bracket denotes $A^{\{\mu\nu\}}=\Xi^{\mu\nu}_{\alpha\beta}A^{\alpha\beta}$. The definitions of the coefficients in Eqs.~\eqref{eq:VAH PL}-\eqref{eq:VAH piperp} can be found in Ref.~\cite{McNelis:2021zji}

\textbf{Initial state.} In our simulations, we fix the initial energy density profile $\epsilon(\tau_0,\xT)\equiv\epsilon_0(\xT)$ as an average energy density profile of an ensemble of Pb+Pb collisions in the 30-40\% centrality class, generated by a saturation-model based initial state generator \cite{Borghini:2022iym}. The same initial condition was also used in Refs.~\cite{Ambrus:2022koq,Ambrus:2022qya}.

The longitudinal pressure $P_L$ is initialized to the early time attractor value of the respective dynamical theory.  At each point $\xT$, the energy-momentum tensor is initialized by
\begin{equation}\label{eq:init}
    T^{\mu\nu}(\tau_0,\xT)=\epsilon_0(\xT) \times \mathrm{diag}\lrp{1,\frac{1}{r+2}, \frac{1}{r+2}, \frac{\tau^{-2} r}{r+2}},
\end{equation}
with $r=P_L/P_\perp\approx 0$ for both \VAH\ and \RTA, because these two models share the free-streaming fixed point at $\tau\to0^+$, while in traditional hydrodynamics, $r$ is slightly negative in this limit \cite{Ambrus:2022koq}. 

In Sec.\,\ref{sec:final-state}, we show the final-state observables from traditional hydrodynamics. Since this description always breaks down at sufficiently early times, a local rescaling manipulation of the initial condition is performed for these hydrodynamic simulations, to properly account for the pre-hydrodynamic evolution. Please see Ref. \cite{Ambrus:2022koq} or Appendix \ref{appendix B} for details.  \\[-0.05in]

Due to the symmetry properties, the evolution of a given initial profile $\tau_0\epsilon_0(\xT)$ depends only on the opacity parameter $\hat \gamma$ \cite{Kurkela:2019kip}, defined by 
\begin{equation}\label{eq:opacity def}
     \hat\gamma = \frac{1}{5\eta/s}\left(\frac{R}{\pi a}\frac{dE^0_\perp}{d\eta}\right)^{1/4}\,.
\end{equation}
Here $a$ is the coefficient in the conformal EoS, $dE_\perp^0/d\eta$ is the initial transverse energy,
 \begin{equation}
 \frac{dE_\perp^0}{d\eta}=\int_{\xT} \tau_0\epsilon_0(\mathbf{x}_\perp),
 \end{equation}
 and the rms transverse radius $R$ that quantifies the transverse size is defined by
 \begin{equation}
     R^2=\left(\frac{dE_\perp^0}{d\eta}\right)^{-1}\int_{\xT} \xT^2 \tau_0\epsilon_0(\mathbf{x}_\perp).
 \end{equation}

By encoding the dependencies on viscosity, energy, and system size, the opacity $\hat\gamma$ provides a universal diluteness criterion, where a small $\hat\gamma$ indicates a dilute system and a large $\hat\gamma$ a dense one. In our discussion, we will vary the opacity via $\eta/s$, as this is the most straightforward way to do so. We stress that this is equivalent to varying the system size or energy scale.

%%%%%%%%%%%%%%%%%%%%%%%%%% 
\subsection{Observables}
%%%%%%%%%%%%%%%%%%%%%%%%%%

%We study the evolution of observables that are particularly sensitive to the degree of hydrodynamic behavior, as they pertain to transverse flow. Some of them technically depend strongly on the initial state geometry, but become almost insensitive to it when normalized to related initial state properties. As we use one fixed initial profile, this normalization is not necessary. Via elliptic flow, we effectively study the elliptic flow response coefficient, and via the final state energy, we effectively study the ratio of transverse energy lost in work performed in the longitudinal expansion. \red{These latter} quantities depend less on initial state properties and are expected to show the same dependence on opacity $\hat{\gamma}$ for all geometries. Therefore, our results allow to infer qualitative conclusions for any collision system.

We study the evolution of observables that are particularly sensitive to the degree of hydrodynamic behavior, as they pertain to transverse flow. All of the observables that we consider can be inferred from the energy-momentum tensor and computed at a fixed time $\tau$. The first observable is the transverse energy $dE_\perp/d\eta$, which characterizes the cooling of the fireball due to longitudinal expansion and is computed as
 \begin{equation}
     \frac{dE_\perp}{d\eta}=\tau\int_{\mathbf{x}_\perp}\left(T^{xx}+T^{yy}\right).
 \end{equation}
Next, to get insights into the thermalization of the system, Sec.~\ref{sec:evolve} also analyzes the evolution of the average inverse Reynolds number, quantifying the deviation from equilibrium,
\begin{equation}
\lra{\mr{Re}^{-1}}_\epsilon = \lra{\lrp{\frac{6\pi^{\mu\nu}\pi_{\mu\nu}}{\epsilon^2}}^{1/2}}_\epsilon,
\end{equation}
where the energy-weighted average over the transverse plane is defined as:
\begin{equation}
\lra{\mc{O}}_\epsilon(\tau)=\frac{\int_{\xT}\mc{O}(\tau,\xT)\epsilon(\tau,\xT)}{\int_{\xT}\epsilon(\tau,\xT)}.
\end{equation}
For \VAH, $\pi^{\mu\nu}$ can be constructed here with Eq.~\eqref{eq:pidcp}. Note that with the initialization (\ref{eq:init})
the initial value of $\lra{\mr{Re}^{-1}}_\epsilon$ approaches 1 as $\tau\to0^+$.

The central observables in our study are those pertaining to the flow response to the initial geometry in the transverse plane. We study radial expansion via the average transverse flow velocity $\lra{u_\perp}_\epsilon$,   
\begin{equation}
\lra{u_\perp}_\epsilon=\lra{\lrp{u_x^2+u_y^2}^{1/2}}_\epsilon,
\end{equation}
and the response of the system to the initial ellipticity $\epsilon_2=\lra{\xT^2e^{2i\phi_{\xT}}}_\epsilon/\lra{\xT^2}_\epsilon$, which is quantified by the momentum anisotropy $\varepsilon_p$,
\begin{equation}
\varepsilon_p=\frac{\int_{\xT}\lrp{T^{xx}-T^{yy}+2iT^{xy}}}{\int_{\xT}\lrp{T^{xx}+T^{yy}}}.
\end{equation}
Although $\varepsilon_p$ is generally complex-valued, its imaginary component vanishes in our simulations due to the alignment of the minor axis of the average initial energy density profile with the $x$-axis. Thus, in Secs.~\ref{sec:evolve} and \ref{sec:final-state} we only consider the real part of $\varepsilon_p$.

We note that the aforementioned observables depend not only on the opacity, but also on the initial state profile $\tau_0\epsilon_0(\xT)$. However, as demonstrated explicitly in Refs.~\cite{Ambrus:2024hks,Ambrus:2024eqa}, the ratios of final state observables to related initial state quantities become almost independent of the initial state geometry, when evaluated at the same opacity. For example, although $\varepsilon_p$ is clearly sensitive to the initial geometry, the ratio $\kappa=\varepsilon_p/\varepsilon_2$ as obtained for different centrality classes in various collision systems follows a common line as a function of $\hat\gamma$. Since in this work, we use a fixed initial profile and vary the opacity $\hat\gamma$ via the shear-viscosity $\eta/s$, a normalization to the initial state eccentricity $\varepsilon_2$ is not necessary, as the latter remains identical. Consequently, when considering momentum anisotropy $\varepsilon_p$, we effectively study the elliptic flow response coefficient $\varepsilon_p/\varepsilon_2$, and when considering the transverse energy $dE_\perp/d\eta$, we effectively study the relative decrease of the transverse energy due to work performed against the longitudinal expansion. Since in accordance with Refs.~\cite{Ambrus:2024hks,Ambrus:2024eqa} these ratios are expected to show the same dependence on opacity $\hat{\gamma}$ for all geometries, our results allow to infer qualitative conclusions for any collision system.

% Some of these observables technically depend strongly on the initial state geometry, but become almost insensitive to it when normalized to related initial state properties\footnote{For example, although $\varepsilon_p$ is sensitive to the initial geometry, its linear response to $\varepsilon_2$, $\kappa=\varepsilon_p/\varepsilon_2$, follows a common line as a function of $\hat\gamma$ after averaging over centrality classes, as shown in Fig.3 of Ref.~\cite{Ambrus:2024hks}}. As we use one fixed initial profile, this normalization is not necessary. Via momentum anisotropy $\varepsilon_p$, we effectively study the elliptic flow response coefficient, and via the transverse energy $dE_\perp/d\eta$, we effectively study the ratio of transverse energy lost in work performed in the longitudinal expansion. \red{These latter} quantities depend less on initial state properties and are expected to show the same dependence on opacity $\hat{\gamma}$ for all geometries. Therefore, our results allow to infer qualitative conclusions for any collision system.}

%FIG,PYY,1
\begin{figure*}[!hbt]\label{fig:evolve}
   \begin{centering}
       \includegraphics[width=0.9\textwidth]{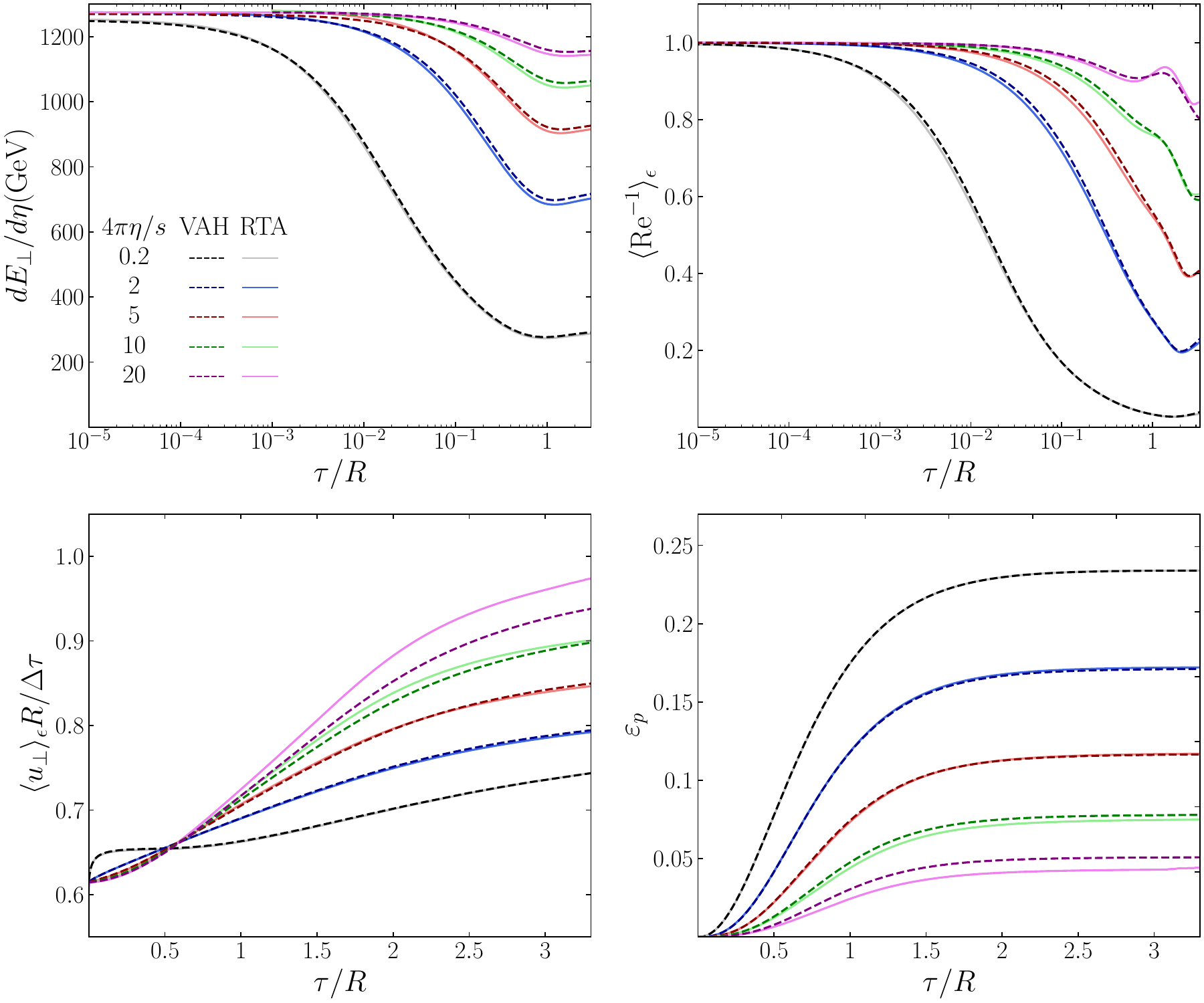}
    \end{centering}
    
    \caption{Time evolution of transverse energy $dE_\perp/d\eta$ (top left), inverse Reynolds number (top right), transverse flow velocity $\langle u_\perp\rangle_\epsilon$ (bottom left) and momentum anisotropy $\varepsilon_p$ (bottom right) from viscous anisotropic hydrodynamics \VAH\ (dashed) and \RTA~(solid) with different specific shear viscosities $\eta/s$.
    }
    \label{fig:evolve}
\end{figure*}

%%%%%%%%%%%%%%%%%%%%%%%%%%%%%%%%%%%
\section{Results}\label{sec:res}
%%%%%%%%%%%%%%%%%%%%%%%%%%%%%%%%%%%
\subsection{Evolution behavior}
\label{sec:evolve}
%%%%%%%%%%%%%%%%%%%%%%%%%%%%%%%%%%%

We now compare the time evolution in \VAH\ and \RTa\ kinetic theory, as shown in Fig.~\ref{fig:evolve}. Keeping in mind the remarks made at the end of Sec.~\ref{sec:models}, varying the shear viscosity allows us to effectively probe the behavior of systems over a wide range in system size.

We start our discussion with the evolution of the transverse energy $dE_\perp/d\eta$, presented in the top left panel. As demonstrated in Ref.~\cite{Ambrus:2021fej}, at times $\tau\ll R$ the dynamics is dominated by a Bjorken-flow-like longitudinal expansion, where the system macroscopically appears as free-streaming (with almost no work done by the longitudinal pressure) and $dE_\perp/d\eta$ remains almost constant. Once the effect of interactions becomes important, longitudinal pressure gradually builds up, and the work done by it causes the transverse energy to decrease, gradually transitioning to a power-law decay $dE_\perp/d\eta\sim\tau^{-1/3}$. After $\tau\simeq R$, the transverse dynamics gradually sets in while the longitudinal expansion rate continues to decrease as $1/\tau$. During this stage, changes in $dE_\perp/d\eta$ are seen to be small. Ultimately, the system approaches the transverse free-streaming limit, where $dE_\perp/d\eta$ is again constant. Lower shear viscosity leads to an earlier build-up of longitudinal pressure and thus to a longer period of loss of transverse energy, yielding a lower final-state plateau. As clearly seen in the figure, \VAH\ accurately reproduces these characteristic features. For $4\pi\eta/s=0.2$, the evolution curve obtained in \VAH\ lies almost on top of those from \RTa\ kinetic theory. For the higher viscosities explored here, the two models still differ by less than 5\%.  

The top right panel shows the evolution of the inverse Reynolds number $\lra{\mr{Re}^{-1}}_\epsilon$. After the initial free-streaming stage in which the inverse Reynolds number stays close to its initial value of 1, $\lra{\mr{Re}^{-1}}_\epsilon$ rapidly decays during hydrodynamization, followed by a slight increase since the strengthening transverse expansion drives the system again away from local equilibrium. While \VAH\ and \RTA~maintain good agreement at moderate viscosities ($4\pi\eta/s\leq 5$), the late-time evolution of $\lra{\mr{Re}^{-1}}_\epsilon$ exhibits larger discrepancies at higher $\eta/s$. They become particularly prominent in late-time evolution at $4\pi\eta/s=20$, where the curves exhibit distinct distortion patterns. 

For both $dE_\perp/d\eta$ and $\lra{\mr{Re}^{-1}}_\epsilon$, we observe slightly higher values in \VAH\ compared to \RTA~during the hydrodynamization process. As this stage is dominated by longitudinal expansion, at each transverse point, the longitudinal dynamics can be approximated by the (0+1)-D evolution along the corresponding local Bjorken attractor. Therefore, we attribute these discrepancies to subtle differences between the \VAH\ and \RTa\ attractors. See Appendix~\ref{appendix A} and Ref.~\cite{Ambrus:2022oji} for more detailed discussions. 

We now examine \VAH's performance in characterizing transverse dynamics. The bottom left panel displays the average transverse flow $\lra{ u_\perp}_\epsilon$, normalized by the ratio of the initial transverse rms radius to the time difference $\Delta\tau=\tau-\tau_0$ in anticipation of a linear growth in time~\cite{Vredevoogd:2008id}. For $4\pi\eta/s=0.2$,  $2$ and $5$, $\lra{u_\perp}_\epsilon$ obtained by \VAH\ and \RTA~are in excellent agreement with each other. However, significant divergences emerge at high viscosities ($4\pi\eta/s\ge10$), where \VAH\ predicts systematically lower transverse flow velocities than \RTa.

Finally, we shift our attention to the bottom right panel, showing the evolution of the momentum anisotropy $\varepsilon_p$. The qualitative behavior is the same in all cases: driven by the anisotropic pressure gradient in the transverse plane, $\varepsilon_p$ continuously increases and eventually saturates at $\tau\simeq 2R$, when the interaction rates become too small to develop further anisotropy. At $4\pi \eta/s=0.2,2$ and $5$, the results of \VAH\ are nearly identical to those of \RTA. However, \VAH\ overestimates the values of $\varepsilon_p$ at high shear viscosities. While \VAH\ maintains $5\%$ accuracy in the final-state value at $4\pi\eta/s=10$, its deviation from \RTA~exceeds $15\%$ at $4\pi\eta/s=20$.

To summarize this section, \VAH\ exhibits excellent agreement with \RTA~when $4\pi\eta/s\le5$; At $4\pi\eta/s=10$, minor discrepancies begin to emerge. At $4\pi\eta/s=20$, while the evolution of the transverse energy $dE_\perp/d\eta$ and the inverse Reynolds number $\lra{\mr{Re}^{-1}}_\epsilon$ is still well described, \VAH\ underestimates the average transverse velocity $\lra{u_\perp}_\epsilon$ and overestimates the momentum anisotropy $\varepsilon_p$. Both features seem to suggest that \VAH\ describes the system to have a slightly higher interaction rate than it actually has, which implies that \VAH\ fails to accurately describe transverse dynamics when the interaction rate is small. However, we emphasize that $4\pi\eta/s=10$ already corresponds to an opacity of $\hat{\gamma}\approx1$, which means that the system is quite dilute. Remarkably, \VAH\ still maintains good agreement with kinetic theory across a wide range of opacity. This point will be further substantiated in the following section through the opacity dependence of final-state observables. 

\subsection{Final-state results}\label{sec:final-state}

\begin{figure}
\includegraphics[width=0.45\textwidth]{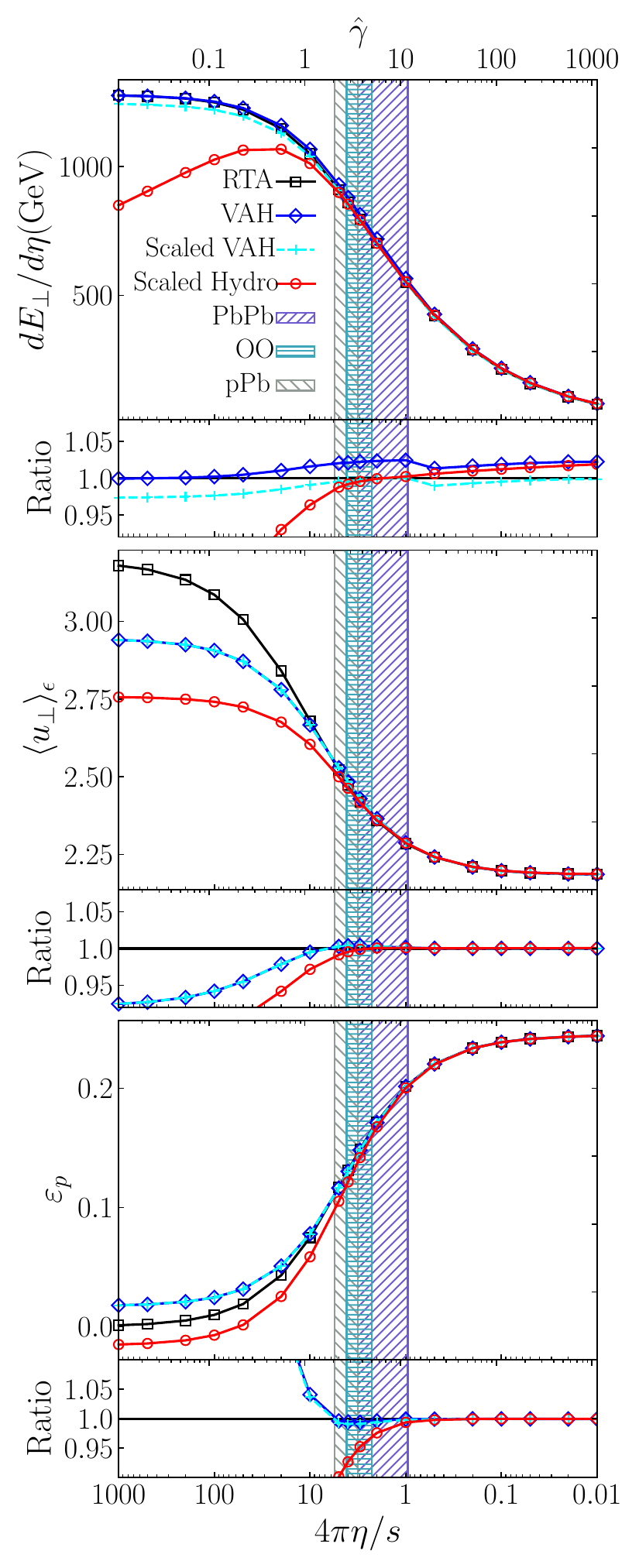}
    \caption{Opacity $\hat\gamma$ or $\eta/s$ dependence of final values of $dE_\perp/d\eta$ (top), $\lra{ u_\perp}_\epsilon$ (middle) and $\varepsilon_p$ from \RTA\ (black), viscous anisotropic hydrodynamics \VAH\ (blue), scaled \VAH\ (dashed cyan) and scaled traditional viscous hydrodynamics \vHLLE\ (red). The bottom subplot in each panel shows ratios to \RTa. Color bands represent opacity  $\hat\gamma$ ranges of Pb+Pb (purple), O+O (blue) and p+Pb(grey) collisions at the LHC.}
    \label{fig:final}
\end{figure}

In this section, we focus on the final state values of transverse energy $dE_\perp/d\eta$, average flow velocity $\langle u_\perp\rangle_\epsilon$, and momentum anisotropy $\varepsilon_p$ extracted at $\tau=3R$ from viscous anisotropic hydrodynamics \VAH, traditional hydrodynamics \vHLLE\ and \RTA. We vary the specific shear viscosity $\eta/s$ from nearly zero to extremely large values, enabling the system to smoothly transition from the free-streaming (opacity $\hat\gamma\to0$) to the ideal fluid (opacity $\hat\gamma\to\infty$) regime. The results are displayed in Fig.~\ref{fig:final}. The \vHLLE\ results with the aforementioned scaling procedure are shown by red curves. For \VAH, both scaled and unscaled results are displayed by cyan and blue curves, respectively. To quantify the deviations of \VAH\ and \vHLLE\ from the \RTA\ benchmark, we also present their results as ratios to \RTA. Additionally, we estimate the opacity of Pb+Pb collisions at $\sqrt{s_{NN}}=2.76\,\mr{TeV}$, O+O collisions at $\sqrt{s_{NN}}=5.36\,\mr{TeV}$ and p+Pb collisions at $\sqrt{s_{NN}}=5.02\,\mr{TeV}$, which are obtained from Table~1 of Ref.~\cite{Ambrus:2024eqa} with the specific shear viscosity adjusted to $\eta/s=0.12$. We mark these opacity regions with different color bands to evaluate the accuracy of \VAH\ and \vHLLE\ in simulating specific collision systems. 

The top panel of Fig.~\ref{fig:final} shows the opacity $\hat\gamma$ dependence of the transverse energy $dE_\perp/d\eta$. We observe that for $\hat\gamma\gtrsim2$, \VAH, scaled \VAH, scaled \vHLLE\ and \RTA~almost perfectly overlap with each other. As the opacity $\hat\gamma$ further decreases, scaled traditional hydrodynamics \vHLLE\ loses the ability to describe the transverse energy $dE_\perp/d\eta$, while \VAH\ remains in excellent agreement with \RTA. Due to the aforementioned slight discrepancy between the attractors of \VAH\ and \RTA, the results of \VAH\ are slightly higher than \RTA---particularly within the typical range of nucleus-nucleus collisions displayed by the colored vertical bands. These minor deviations can be mitigated using our initial state scaling approach, as shown by the scaled \VAH\ curves. For \VAH, this scaling is equivalent to multiplying the initial energy density by a global constant. Consequently, for the dimensionless observables $\lra{u_\perp}_\epsilon$ and $\varepsilon_p$, it introduces an opacity shift of only $\sim 10^{-3}$, rendering its effect virtually negligible.

We then turn our attention to the average transverse flow velocity $\langle u_\perp\rangle_\epsilon$. The value of $\lra{u_\perp}_\epsilon$ decreases as the system becomes denser, approaching two distinct limits in the $\hat\gamma\to0$ and $\hat\gamma\to\infty$ regimes. This behavior is qualitatively reproduced by both \VAH\ and \SVH. For $\hat\gamma\gtrsim 2$,  both \VAH\ and \SVH~yield nearly identical results to \RTa. At lower opacity, $\lra{u_\perp}_\epsilon$ of \VAH\ lies between \RTA~and \SVH, demonstrating its superior performance over traditional hydrodynamics in describing radial expansion. 
The hierarchy $\lra{u_\perp}_\epsilon(\text{\RTa}) > \lra{u_\perp}_\epsilon(\text{\VAH}) > \lra{u_\perp}_\epsilon(\text{\SVH})$ indicates that \VAH\ is more dissipative than \RTA\ and scaled traditional hydrodynamics \vHLLE\ is more dissipative than \VAH, leading to less transverse flow, especially in the low opacity limit.

Finally, we analyze the results of momentum anisotropy $\varepsilon_p$. As the system becomes increasingly dense, $\varepsilon_p$ continues to rise until saturating at the ideal hydro limit---a value identical across all three models. As the opacity drops to the characteristic opacity range of O+O and p+Pb collisions, \SVH~calculations gradually deviate from \RTA~(when $\hat{\gamma} \lesssim 3$), whereas \VAH\ maintains remarkably close agreement, down to $\hat{\gamma} \simeq 1$. This underscores the superior capability of \VAH\ in modeling collective flow signals in small collision systems. At lower opacity, the behavior of $\varepsilon_p$ obtained in these three models differ significantly. In \RTA, the value of $\varepsilon_p$ converges to zero in the free-streaming limit ($\hat{\gamma} \to 0$), while \VAH\ and \SVH~yield positive and negative $\varepsilon_p$, respectively. This sign discrepancy arises because the dissipative terms in each model contribute differently to the momentum anisotropy $\varepsilon_p$. See Appendix \ref{appendix B} for further discussion.

\begin{figure*}[!t]
\centering
\begin{tabular}{cc}
\includegraphics[width=0.45\textwidth]{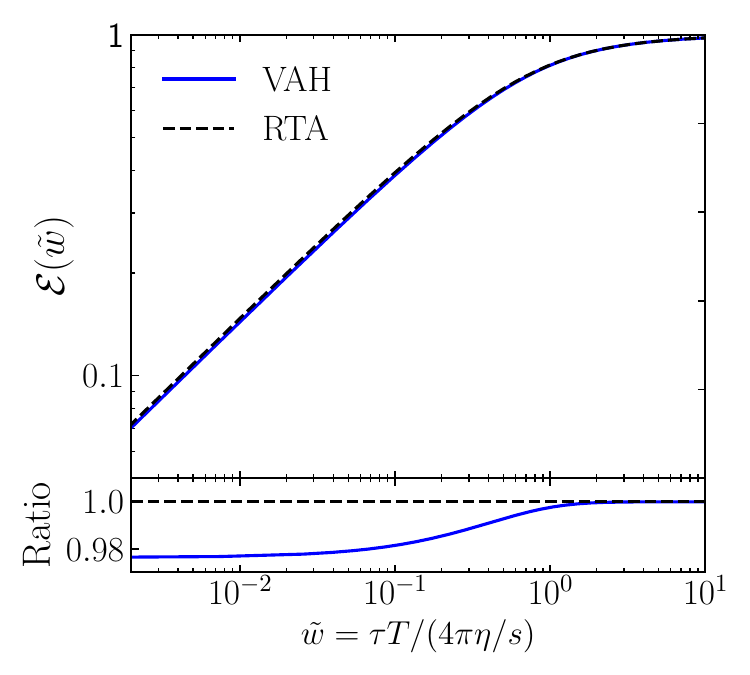} &
\includegraphics[width=0.443\textwidth]{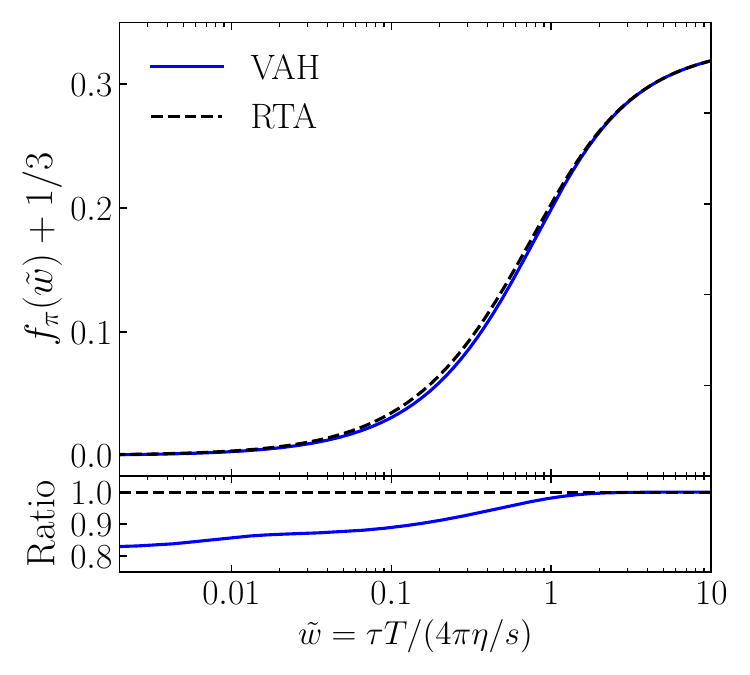}\\
\end{tabular}
\caption{$\mc{E}(\tilde w)$ and $f_\pi(\tilde w)+1/3$ from \VAH~ (blue) and \RTA~(dashed black). The ratios of \VAH\ to \RTA~are plotted in the subplots below the main panels.}
\label{fig:attractor}
\end{figure*}

%%%%%%%%%%%%%%%%%%%%%%%%%%%%%%%%%%%%%%
\section{Conclusions}\label{sec:conc}
%%%%%%%%%%%%%%%%%%%%%%%%%%%%%%%%%%%%%%

In this work, we systematically compared the evolution of viscous anisotropic hydrodynamics \VAH\  and traditional viscous hydrodynamics \vHLLE\  with \RTA~ for a conformal system undergoing $(2{+}1)$-dimensional expansion with longitudinal boost-invariance. Earlier work had established that for (0+1)-D Bjorken flow (longitudinal boost-invariant expansion without transverse flow), viscous anisotropic hydrodynamics describes the evolution of the macroscopic hydrodynamic fields associated with the underlying \RTA~almost perfectly \cite{Florkowski:2013lya, Bazow:2013ifa}. A key goal of the present study was to explore to what extent transverse flow effects, as they occur in more realistic systems with spatially anisotropic initial densities of finite transverse extent, degrade the precision of viscous anisotropic hydrodynamics as a macroscopic effective description of the underlying microscopic kinetic evolution. To quantify the differences between the various descriptions, we compare their predictions for the transverse energy $dE_\perp/d\eta$, average inverse Reynolds number $\lra{\mr{Re}^{-1}}_\epsilon$, average transverse flow velocity $\lra{u_\perp}_\epsilon$, and momentum anisotropy $\varepsilon_p$ as functions of time $\tau/R$ and opacity $\hat{\gamma}$. Surprisingly good agreement between \VAH\ and \RTA~is observed during the far-off-equilibrium earliest evolution stage ($\tau{\,\ll\,}R$) when the transverse expansion is weak; minor deviations arise only from the slight differences between the Bjorken attractors of these two models. Significant discrepancies emerge in the late-time evolution when transverse expansion dominates; they are limited, however, to cases with exceptionally large shear viscosity ($4\pi\eta/s\gtrsim10$), corresponding to low opacity ($\hat\gamma\lesssim1$). By analyzing the final-state values of observables across the full range of opacity $\hat\gamma$, we demonstrate that in smaller systems (such as those created in O+O or p+Pb collisions), traditional hydrodynamics gradually fails, showing increasing deviations from the underlying microscopic theory, while viscous anisotropic hydrodynamics \VAH\ maintains excellent agreement throughout. Qualitatively, our observations can be summarized by the simple statement that, for a fixed opacity of the medium, among the three models studied, \RTA\ exhibits the least amount of dissipation, \VAH\ is slightly more dissipative, and traditional viscous hydrodynamics features the largest dissipative effects, especially at small opacities. 

In summary, we conclude that when compared to traditional viscous hydrodynamics, viscous anisotropic hydrodynamics significantly expands the regime of applicability of the macroscopic hydrodynamic approach, making it particularly promising for an improved macroscopic description of the evolution of small systems.

\begin{acknowledgments}
The authors thank Baochi Fu, Shujun Zhao and Yili Wang for fruitful discussions. Y.P. and H.S. are supported by the National Science Foundation of China under Grant No.12575138 and No.12247107.   C.W. has received funding from the European Research Council (ERC) under the European Union’s Horizon 2020 research and innovation programme (grant number: 101089093 / project acronym: High-TheQ). 
V.E.A. gratefully acknowledges funding by the EU’s NextGenerationEU instrument through the National Recovery and Resilience Plan of Romania - Pillar III-C9-I8, managed by the Ministry of Research, Innovation and Digitization, within the project entitled ``Facets of Rotating Quark-Gluon Plasma'' (FORQ), contract no.~760079/23.05.2023 code CF 103/15.11.2022. 
VEA, CW and SS gratefully acknowledge  support by the Deutsche Forschungsgemeinschaft (DFG, German Research Foundation) through the CRC-TR 211 ``Strong-interaction matter under extreme conditions,'' Project No. 315477589--TRR 211.
Views and opinions expressed are however those of the authors only and do not necessarily reflect those of the European Union or the European Research Council. Neither the European Union nor the granting authority can be held responsible for them.
\end{acknowledgments}

\section*{Data availability}

The data that support the findings of this article are openly available \cite{peng_2025_Zenodo}.

\appendix

\section{Universal Bjorken Attractor}\label{appendix A}

We now discuss how the slight early time deviation of viscous anisotropic hydrodynamics \VAH\ from \RTA~can be explained via the Bjorken flow attractor curves corresponding to these two theories. In $(0{+}1)$-D Bjorken flow, the energy-momentum tensor is diagonal,
\begin{equation}
    T^{\mu\nu} = \mr{diag}\lrp{\epsilon, P_\perp, P_\perp, \tau^{-2}P_L}.
\end{equation}
Under the assumption of constant shear viscosity and a conformal equation of state, the universal attractor solutions for pressure anisotropy $f_\pi=P_L/\epsilon-1/3$ and scaled energy density $\mc{E}$, which for a given initial energy density can be expressed as~\cite{Giacalone:2019ldn}
\begin{equation}\label{eq:attractor e}
\frac{\mc{E}(\tilde w)}{\mc{E}(\tilde w_0)}=\frac{\tau^{4/3}\epsilon(\tau)}{\tau_0^{4/3}\epsilon(\tau_0)},
\end{equation}
are functions only of the conformal scaling parameter $\tilde w$ defined by
\begin{equation}
    \tilde w=\frac{5\tau}{4\pi\tau_R(\tau)}=\frac{\tau T(\tau)}{4\pi\eta/s}.
\end{equation}

\begin{table}[t]
  \centering
  \begin{tabular}{@{\extracolsep{1cm}}lcc}
    \hline
    \textbf{Models} & $C_{\infty}$ & $\gamma$ \\
    \hline
    RTA    & 0.88 & 4/9 \\
    VAH   & 0.90 & 4/9 \\
    Hydro  & 0.82 & 0.526 \\
    \hline
  \end{tabular}
  \caption{ $C_{\infty}$ and $\gamma$ of \RTA, viscous anisotropic hydrodynamics \VAH\ and traditional viscous hydrodynamics \vHLLE.}
  \label{tab:models}
\end{table}

The universal function $\mc{E}(\tilde w)$ exhibits the following asymptotic behavior at early ($\tilde w\ll1$) and late ($\tilde w\gg 1$) times:
\begin{equation}\label{eq:asymptotic e}
    \mc{E}(\tilde w\ll1)=C_{\infty}^{-1}\tilde w^\gamma,\quad\mc{E}(\tilde w\gg 1)=1-\frac1{4\pi\tilde w}.
\end{equation}
The normalization constant $C_{\infty}$ and early time power law exponent $\gamma$ of \RTA, \VAH\ and  \vHLLE\ are listed in Table~\ref{tab:models}. 

We now briefly review the scaling method used in our simulations. The aim is to counteract the differences in the pre-equilibrium evolution in the hydrodynamic theories compared to \RTA. For a hydrodynamic theory with given parameters $C_{\infty}$ and $\gamma$, we require that under the Bjorken flow, the late-stage evolution of the energy density converges to that in \RTA. Based on Eqs.~\eqref{eq:attractor e} and \eqref{eq:asymptotic e}, the initial energy density in such a theory must be scaled by~\cite{Ambrus:2022koq}
{
\begin{multline}
    \epsilon_{0}^{\mr{scaled}}(\xT)\\
    =\left[\lrp{\frac{4\pi\eta/s}{\tau_0}a^{1/4}}^{\frac12-\frac{9\gamma}{8}}\lrp{\frac{C_\infty^{\mr{RTA}}}{C_{\infty}}}^{9/8}\epsilon_0(\xT)\right]^{\frac{8/9}{1-\gamma/4}}.
\end{multline}
}
Fig.~\ref{fig:attractor} shows $\mc{E}(\tilde w)$ and $f_\pi(\tilde w)$ from \VAH\ and \RTA. These results explain the discrepancies of $dE_\perp/d\eta$ and $\lra{\mr{Re}^{-1}}_\epsilon$ during the pre-equilibrium stage, which are shown in Fig.~\ref{fig:evolve}. Specifically, in the early evolution stage, both $\mc{E}(\tilde w)$ and $f_\pi(\tilde w)$ of VAH are slightly smaller than in \RTA, whereas at $\tilde w\sim 1$ the attractors of these two models converge to each other. As
\begin{align}
    f_\pi(\tilde{w})=\frac{2}{3}\left[\frac{1}{4}-\left(\frac{d\ln \mc{E}(\tilde{w})}{d \ln \tilde{w}}\right)^{-1}\right]^{-1}
\end{align}
is fully determined by $\mc{E}(\tilde{w})$~\cite{Ambrus:2022koq} and the normalization of $\mc{E}(\tilde{w})$ is fixed by the late time limit, while the early time free-streaming limit is the same in both theories,
we can conclude that the difference between the curves comes from deviations in the dynamics on the intermediate timescale $w\sim1$, so from the details of the hydrodynamization process.

\begin{figure}[!t]
 \includegraphics[width=0.45\textwidth]{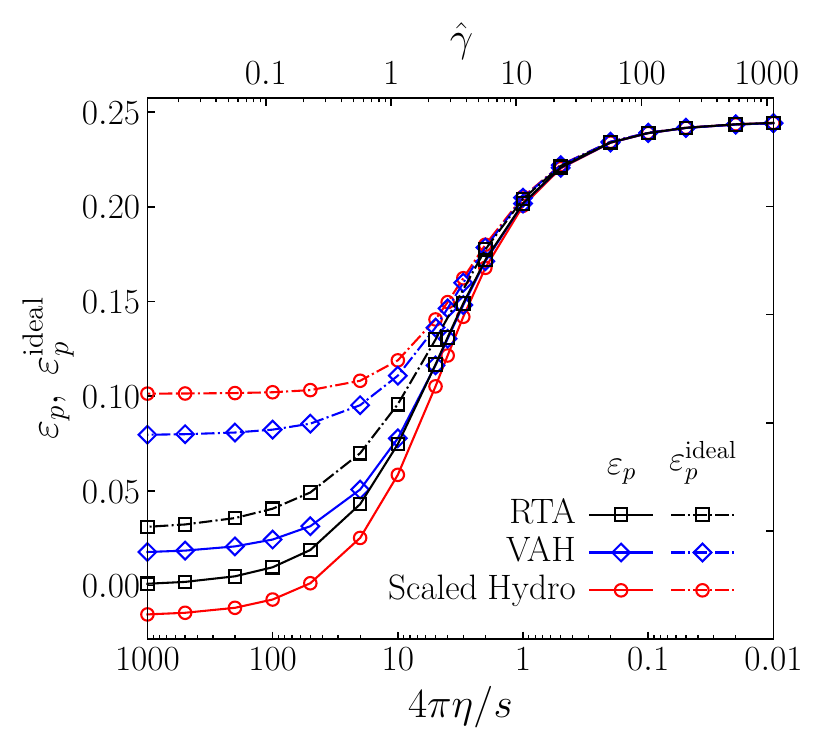}
 \caption{Opacity $\hat\gamma$ or $\eta/s$ dependence of the final values of total elliptic flow $\varepsilon_p$ (solid line) and elliptic flow $\varepsilon_p^{\idl}$ from the ideal part of the energy-momentum tensor (dot-dashed line) from \RTA\ (black), \VAH\ (blue) and \SVH~(red).
 }
 \label{fig:epid}
\end{figure}

In (2+1)-D simulations, the longitudinal expansion dominates until around $\tau\sim R$, thus the transverse energy can be approximated for early times as
\begin{equation}
    \frac{dE_\perp}{d\eta} \approx \int_{\xT}\lrp{\frac{2}{3}-f_\pi[\tilde w(\xT)]}\epsilon(\tau,\xT),
\end{equation}
where $\tilde w(\xT)$ and $\epsilon(\tau,\xT)$ are determined locally. Therefore, the transverse energy of \VAH\ becomes slightly larger due to a smaller denominator on the left-hand side of Eq.~\eqref{eq:attractor e}. Qualitatively, the overestimation of average inverse Reynolds number in \VAH\ is straightforward to understand, since for $\tau\lesssim R$ the inverse Reynolds number can be approximately expressed as $\mr{Re}^{-1}\approx -3f_\pi$.

\section{Dissipative effects on the momentum anisotropy}\label{appendix B}

To further elucidate why the different behaviors of the dissipative terms corresponding to the shear-stress tensor in \RTA, \VAH\ and \SVH~lead to different signs of the momentum anisotropy $\varepsilon_p$, we analyze separately the contribution due to the ideal part of the energy-momentum tensor, 
\begin{equation}\label{eq:epid}
    \varepsilon_p^{\mathrm{ideal}}\equiv \frac{\int d^2\xT\left(T^{xx}_{\idl}-T^{yy}_{\idl}+2iT^{xy}_{\idl}\right)}{\int d^2\xT\left(T^{xx}_{\idl}+T^{yy}_{\idl}\right)},
\end{equation}
where $T^{\mu\nu}_{\idl}= \epsilon u^\mu u^\nu - P\Delta^{\mu\nu}$. The results are presented in Fig.~\ref{fig:epid}. For all three models, $\varepsilon_p^{\mr{ideal}}$ converges to $\varepsilon_p$ in the large-opacity region due to the negligible dissipation effects. As opacity decreases, $\varepsilon_p^{\mr{ideal}}$ becomes larger than $\varepsilon_p$. Notably, both \VAH\ and \SVH~ overestimate $\varepsilon_p^{\mr{ideal}}$ compared to \RTA, with the hierarchy $\varepsilon_p^{\mr{ideal}}(\text{\SVH})>\varepsilon_p^{\mr{ideal}}(\text{\VAH})>\varepsilon_p^{\mr{ideal}}(\text{\RTa})$.    
%{\color{purple} 
%Notably, the difference between the two $\varepsilon_p^{\idl}$-curves strongly correlates with the difference of the two $\varepsilon_p$-curves.
%[VEA: anti-correlates?]
%}
We conclude that the dissipative contribution 
inhibits the anisotropic flow created due to the ideal part of the energy-momentum tensor. In the \RTA, as $\hat\gamma\to0$, the dissipative term completely cancels $\varepsilon_p^{\mr{ideal}}$, driving $\varepsilon_p$ to zero.  \VAH\ creates a smaller $\varepsilon_p^{\idl}$ compared to \SVH, yet exhibits a weaker dissipative suppression effect. The difference in sign between the total $\varepsilon_p$ obtained in \VAH\ and \SVH~ is a curious consequence of the fact that the larger $\varepsilon_p^{\idl}$ generated in traditional hydrodynamics is significantly more heavily suppressed by the dissipative terms than in \VAH, leading to a negative total $\varepsilon_p$.
As a consequence, the (negative) dissipative contribution, $\varepsilon^{\rm diss}_p \equiv \varepsilon_p - \varepsilon^{\rm ideal}_p$, obeys an inverted hierarchy, $\varepsilon_p^{\mr{diss}}(\text{\SVH}) < \varepsilon_p^{\mr{diss}}(\text{\VAH}) < \varepsilon_p^{\mr{diss}}(\text{\RTa})$, with Scaled Hydro  being more dissipative than \VAH, which is more dissipative than \RTA. This conclusion is consistent with the $\lra{u_\perp}_\epsilon$ ordering shown in Fig.~\ref{fig:final}.

\bibliography{exp,kinetic,hydro_vah}

\begin{thebibliography}{98}%
\makeatletter
\providecommand \@ifxundefined [1]{%
 \@ifx{#1\undefined}
}%
\providecommand \@ifnum [1]{%
 \ifnum #1\expandafter \@firstoftwo
 \else \expandafter \@secondoftwo
 \fi
}%
\providecommand \@ifx [1]{%
 \ifx #1\expandafter \@firstoftwo
 \else \expandafter \@secondoftwo
 \fi
}%
\providecommand \natexlab [1]{#1}%
\providecommand \enquote  [1]{``#1''}%
\providecommand \bibnamefont  [1]{#1}%
\providecommand \bibfnamefont [1]{#1}%
\providecommand \citenamefont [1]{#1}%
\providecommand \href@noop [0]{\@secondoftwo}%
\providecommand \href [0]{\begingroup \@sanitize@url \@href}%
\providecommand \@href[1]{\@@startlink{#1}\@@href}%
\providecommand \@@href[1]{\endgroup#1\@@endlink}%
\providecommand \@sanitize@url [0]{\catcode `\\12\catcode `\$12\catcode
  `\&12\catcode `\#12\catcode `\^12\catcode `\_12\catcode `\%12\relax}%
\providecommand \@@startlink[1]{}%
\providecommand \@@endlink[0]{}%
\providecommand \url  [0]{\begingroup\@sanitize@url \@url }%
\providecommand \@url [1]{\endgroup\@href {#1}{\urlprefix }}%
\providecommand \urlprefix  [0]{URL }%
\providecommand \Eprint [0]{\href }%
\providecommand \doibase [0]{http://dx.doi.org/}%
\providecommand \selectlanguage [0]{\@gobble}%
\providecommand \bibinfo  [0]{\@secondoftwo}%
\providecommand \bibfield  [0]{\@secondoftwo}%
\providecommand \translation [1]{[#1]}%
\providecommand \BibitemOpen [0]{}%
\providecommand \bibitemStop [0]{}%
\providecommand \bibitemNoStop [0]{.\EOS\space}%
\providecommand \EOS [0]{\spacefactor3000\relax}%
\providecommand \BibitemShut  [1]{\csname bibitem#1\endcsname}%
\let\auto@bib@innerbib\@empty
%</preamble>
\bibitem [{\citenamefont {Kolb}\ and\ \citenamefont
  {Heinz}(2003)}]{Kolb:2003dz}%
  \BibitemOpen
  \bibfield  {author} {\bibinfo {author} {\bibfnamefont {P.~F.}\ \bibnamefont
  {Kolb}}\ and\ \bibinfo {author} {\bibfnamefont {U.}~\bibnamefont {Heinz}},\
  }\enquote {\bibinfo {title} {{Hydrodynamic description of ultrarelativistic
  heavy ion collisions}},}\ in\ \href {\doibase 10.1142/9789812795533_0010}
  {\emph {\bibinfo {booktitle} {{Quark-Gluon Plasma 3}}}},\ \bibinfo {editor}
  {edited by\ \bibinfo {editor} {\bibfnamefont {R.~C.}\ \bibnamefont {Hwa}}\
  and\ \bibinfo {editor} {\bibfnamefont {X.-N.}\ \bibnamefont {Wang}}}\
  (\bibinfo {year} {2003})\ pp.\ \bibinfo {pages} {634--714},\ \Eprint
  {http://arxiv.org/abs/nucl-th/0305084} {arXiv:nucl-th/0305084 [nucl-th]}
  \BibitemShut {NoStop}%
\bibitem [{\citenamefont {Teaney}(2010)}]{Teaney:2009qa}%
  \BibitemOpen
  \bibfield  {author} {\bibinfo {author} {\bibfnamefont {D.~A.}\ \bibnamefont
  {Teaney}},\ }\enquote {\bibinfo {title} {{Viscous Hydrodynamics and the Quark
  Gluon Plasma}},}\ in\ \href {\doibase 10.1142/9789814293297_0004} {\emph
  {\bibinfo {booktitle} {{Quark-Gluon Plasma 4}}}},\ \bibinfo {editor} {edited
  by\ \bibinfo {editor} {\bibfnamefont {R.~C.}\ \bibnamefont {Hwa}}\ and\
  \bibinfo {editor} {\bibfnamefont {X.-N.}\ \bibnamefont {Wang}}}\ (\bibinfo
  {year} {2010})\ pp.\ \bibinfo {pages} {207--266},\ \Eprint
  {http://arxiv.org/abs/0905.2433} {arXiv:0905.2433 [nucl-th]} \BibitemShut
  {NoStop}%
\bibitem [{\citenamefont {Heinz}\ and\ \citenamefont
  {Snellings}(2013)}]{Heinz:2013th}%
  \BibitemOpen
  \bibfield  {author} {\bibinfo {author} {\bibfnamefont {U.}~\bibnamefont
  {Heinz}}\ and\ \bibinfo {author} {\bibfnamefont {R.}~\bibnamefont
  {Snellings}},\ }\href {\doibase 10.1146/annurev-nucl-102212-170540}
  {\bibfield  {journal} {\bibinfo  {journal} {Ann. Rev. Nucl. Part. Sci.}\
  }\textbf {\bibinfo {volume} {63}},\ \bibinfo {pages} {123} (\bibinfo {year}
  {2013})},\ \Eprint {http://arxiv.org/abs/1301.2826} {arXiv:1301.2826
  [nucl-th]} \BibitemShut {NoStop}%
\bibitem [{\citenamefont {Hirano}\ \emph {et~al.}(2013)\citenamefont {Hirano},
  \citenamefont {Huovinen}, \citenamefont {Murase},\ and\ \citenamefont
  {Nara}}]{Hirano:2012kj}%
  \BibitemOpen
  \bibfield  {author} {\bibinfo {author} {\bibfnamefont {T.}~\bibnamefont
  {Hirano}}, \bibinfo {author} {\bibfnamefont {P.}~\bibnamefont {Huovinen}},
  \bibinfo {author} {\bibfnamefont {K.}~\bibnamefont {Murase}}, \ and\ \bibinfo
  {author} {\bibfnamefont {Y.}~\bibnamefont {Nara}},\ }\href {\doibase
  10.1016/j.ppnp.2013.02.002} {\bibfield  {journal} {\bibinfo  {journal} {Prog.
  Part. Nucl. Phys.}\ }\textbf {\bibinfo {volume} {70}},\ \bibinfo {pages}
  {108} (\bibinfo {year} {2013})},\ \Eprint {http://arxiv.org/abs/1204.5814}
  {arXiv:1204.5814 [nucl-th]} \BibitemShut {NoStop}%
\bibitem [{\citenamefont {Florkowski}\ \emph {et~al.}(2018)\citenamefont
  {Florkowski}, \citenamefont {Heller},\ and\ \citenamefont
  {Spalinski}}]{Florkowski:2017olj}%
  \BibitemOpen
  \bibfield  {author} {\bibinfo {author} {\bibfnamefont {W.}~\bibnamefont
  {Florkowski}}, \bibinfo {author} {\bibfnamefont {M.~P.}\ \bibnamefont
  {Heller}}, \ and\ \bibinfo {author} {\bibfnamefont {M.}~\bibnamefont
  {Spalinski}},\ }\href {\doibase 10.1088/1361-6633/aaa091} {\bibfield
  {journal} {\bibinfo  {journal} {Rept. Prog. Phys.}\ }\textbf {\bibinfo
  {volume} {81}},\ \bibinfo {pages} {046001} (\bibinfo {year} {2018})},\
  \Eprint {http://arxiv.org/abs/1707.02282} {arXiv:1707.02282 [hep-ph]}
  \BibitemShut {NoStop}%
\bibitem [{\citenamefont {Song}\ \emph {et~al.}(2017)\citenamefont {Song},
  \citenamefont {Zhou},\ and\ \citenamefont {Gajdosova}}]{Song:2017wtw}%
  \BibitemOpen
  \bibfield  {author} {\bibinfo {author} {\bibfnamefont {H.}~\bibnamefont
  {Song}}, \bibinfo {author} {\bibfnamefont {Y.}~\bibnamefont {Zhou}}, \ and\
  \bibinfo {author} {\bibfnamefont {K.}~\bibnamefont {Gajdosova}},\ }\href
  {\doibase 10.1007/s41365-017-0245-4} {\bibfield  {journal} {\bibinfo
  {journal} {Nucl. Sci. Tech.}\ }\textbf {\bibinfo {volume} {28}},\ \bibinfo
  {pages} {99} (\bibinfo {year} {2017})},\ \Eprint
  {http://arxiv.org/abs/1703.00670} {arXiv:1703.00670 [nucl-th]} \BibitemShut
  {NoStop}%
\bibitem [{\citenamefont {Heinz}\ and\ \citenamefont
  {Schenke}(2024)}]{Heinz:2024jwu}%
  \BibitemOpen
  \bibfield  {author} {\bibinfo {author} {\bibfnamefont {U.}~\bibnamefont
  {Heinz}}\ and\ \bibinfo {author} {\bibfnamefont {B.}~\bibnamefont {Schenke}}\
  }(\bibinfo {year} {2024})\ \Eprint {http://arxiv.org/abs/2412.19393}
  {arXiv:2412.19393 [nucl-th]} \BibitemShut {NoStop}%
\bibitem [{\citenamefont {Aidala}\ \emph {et~al.}(2018)\citenamefont {Aidala}
  \emph {et~al.}}]{PHENIX:2017xrm}%
  \BibitemOpen
  \bibfield  {author} {\bibinfo {author} {\bibfnamefont {C.}~\bibnamefont
  {Aidala}} \emph {et~al.} (\bibinfo {collaboration} {PHENIX}),\ }\href
  {\doibase 10.1103/PhysRevLett.120.062302} {\bibfield  {journal} {\bibinfo
  {journal} {Phys. Rev. Lett.}\ }\textbf {\bibinfo {volume} {120}},\ \bibinfo
  {pages} {062302} (\bibinfo {year} {2018})},\ \Eprint
  {http://arxiv.org/abs/1707.06108} {arXiv:1707.06108 [nucl-ex]} \BibitemShut
  {NoStop}%
\bibitem [{\citenamefont {Aidala}\ \emph {et~al.}(2019)\citenamefont {Aidala}
  \emph {et~al.}}]{PHENIX:2018lia}%
  \BibitemOpen
  \bibfield  {author} {\bibinfo {author} {\bibfnamefont {C.}~\bibnamefont
  {Aidala}} \emph {et~al.} (\bibinfo {collaboration} {PHENIX}),\ }\href
  {\doibase 10.1038/s41567-018-0360-0} {\bibfield  {journal} {\bibinfo
  {journal} {Nature Phys.}\ }\textbf {\bibinfo {volume} {15}},\ \bibinfo
  {pages} {214} (\bibinfo {year} {2019})},\ \Eprint
  {http://arxiv.org/abs/1805.02973} {arXiv:1805.02973 [nucl-ex]} \BibitemShut
  {NoStop}%
\bibitem [{\citenamefont {Abelev}\ \emph {et~al.}(2014)\citenamefont {Abelev}
  \emph {et~al.}}]{ALICE:2014dwt}%
  \BibitemOpen
  \bibfield  {author} {\bibinfo {author} {\bibfnamefont {B.~B.}\ \bibnamefont
  {Abelev}} \emph {et~al.} (\bibinfo {collaboration} {ALICE}),\ }\href
  {\doibase 10.1103/PhysRevC.90.054901} {\bibfield  {journal} {\bibinfo
  {journal} {Phys. Rev. C}\ }\textbf {\bibinfo {volume} {90}},\ \bibinfo
  {pages} {054901} (\bibinfo {year} {2014})},\ \Eprint
  {http://arxiv.org/abs/1406.2474} {arXiv:1406.2474 [nucl-ex]} \BibitemShut
  {NoStop}%
\bibitem [{\citenamefont {Aaboud}\ \emph {et~al.}(2017)\citenamefont {Aaboud}
  \emph {et~al.}}]{ATLAS:2017hap}%
  \BibitemOpen
  \bibfield  {author} {\bibinfo {author} {\bibfnamefont {M.}~\bibnamefont
  {Aaboud}} \emph {et~al.} (\bibinfo {collaboration} {ATLAS}),\ }\href
  {\doibase 10.1140/epjc/s10052-017-4988-1} {\bibfield  {journal} {\bibinfo
  {journal} {Eur. Phys. J. C}\ }\textbf {\bibinfo {volume} {77}},\ \bibinfo
  {pages} {428} (\bibinfo {year} {2017})},\ \Eprint
  {http://arxiv.org/abs/1705.04176} {arXiv:1705.04176 [hep-ex]} \BibitemShut
  {NoStop}%
\bibitem [{\citenamefont {Acharya}\ \emph {et~al.}(2019)\citenamefont {Acharya}
  \emph {et~al.}}]{ALICE:2019zfl}%
  \BibitemOpen
  \bibfield  {author} {\bibinfo {author} {\bibfnamefont {S.}~\bibnamefont
  {Acharya}} \emph {et~al.} (\bibinfo {collaboration} {ALICE}),\ }\href
  {\doibase 10.1103/PhysRevLett.123.142301} {\bibfield  {journal} {\bibinfo
  {journal} {Phys. Rev. Lett.}\ }\textbf {\bibinfo {volume} {123}},\ \bibinfo
  {pages} {142301} (\bibinfo {year} {2019})},\ \Eprint
  {http://arxiv.org/abs/1903.01790} {arXiv:1903.01790 [nucl-ex]} \BibitemShut
  {NoStop}%
\bibitem [{\citenamefont {Khachatryan}\ \emph {et~al.}(2017)\citenamefont
  {Khachatryan} \emph {et~al.}}]{CMS:2016fnw}%
  \BibitemOpen
  \bibfield  {author} {\bibinfo {author} {\bibfnamefont {V.}~\bibnamefont
  {Khachatryan}} \emph {et~al.} (\bibinfo {collaboration} {CMS}),\ }\href
  {\doibase 10.1016/j.physletb.2016.12.009} {\bibfield  {journal} {\bibinfo
  {journal} {Phys. Lett. B}\ }\textbf {\bibinfo {volume} {765}},\ \bibinfo
  {pages} {193} (\bibinfo {year} {2017})},\ \Eprint
  {http://arxiv.org/abs/1606.06198} {arXiv:1606.06198 [nucl-ex]} \BibitemShut
  {NoStop}%
\bibitem [{\citenamefont {Sirunyan}\ \emph {et~al.}(2018)\citenamefont
  {Sirunyan} \emph {et~al.}}]{CMS:2017kcs}%
  \BibitemOpen
  \bibfield  {author} {\bibinfo {author} {\bibfnamefont {A.~M.}\ \bibnamefont
  {Sirunyan}} \emph {et~al.} (\bibinfo {collaboration} {CMS}),\ }\href
  {\doibase 10.1103/PhysRevLett.120.092301} {\bibfield  {journal} {\bibinfo
  {journal} {Phys. Rev. Lett.}\ }\textbf {\bibinfo {volume} {120}},\ \bibinfo
  {pages} {092301} (\bibinfo {year} {2018})},\ \Eprint
  {http://arxiv.org/abs/1709.09189} {arXiv:1709.09189 [nucl-ex]} \BibitemShut
  {NoStop}%
\bibitem [{\citenamefont {M{\"a}ntysaari}\ \emph {et~al.}(2025)\citenamefont
  {M{\"a}ntysaari}, \citenamefont {Schenke}, \citenamefont {Shen},\ and\
  \citenamefont {Zhao}}]{Mantysaari:2025tcg}%
  \BibitemOpen
  \bibfield  {author} {\bibinfo {author} {\bibfnamefont {H.}~\bibnamefont
  {M{\"a}ntysaari}}, \bibinfo {author} {\bibfnamefont {B.}~\bibnamefont
  {Schenke}}, \bibinfo {author} {\bibfnamefont {C.}~\bibnamefont {Shen}}, \
  and\ \bibinfo {author} {\bibfnamefont {W.}~\bibnamefont {Zhao}},\ }\href
  {\doibase 10.1103/gf4y-p5j7} {\bibfield  {journal} {\bibinfo  {journal}
  {Phys. Rev. Lett.}\ }\textbf {\bibinfo {volume} {135}},\ \bibinfo {pages}
  {022302} (\bibinfo {year} {2025})},\ \Eprint
  {http://arxiv.org/abs/2502.05138} {arXiv:2502.05138 [nucl-th]} \BibitemShut
  {NoStop}%
\bibitem [{\citenamefont {Shen}\ \emph {et~al.}(2017)\citenamefont {Shen},
  \citenamefont {Paquet}, \citenamefont {Denicol}, \citenamefont {Jeon},\ and\
  \citenamefont {Gale}}]{Shen:2016zpp}%
  \BibitemOpen
  \bibfield  {author} {\bibinfo {author} {\bibfnamefont {C.}~\bibnamefont
  {Shen}}, \bibinfo {author} {\bibfnamefont {J.-F.}\ \bibnamefont {Paquet}},
  \bibinfo {author} {\bibfnamefont {G.~S.}\ \bibnamefont {Denicol}}, \bibinfo
  {author} {\bibfnamefont {S.}~\bibnamefont {Jeon}}, \ and\ \bibinfo {author}
  {\bibfnamefont {C.}~\bibnamefont {Gale}},\ }\href {\doibase
  10.1103/PhysRevC.95.014906} {\bibfield  {journal} {\bibinfo  {journal} {Phys.
  Rev. C}\ }\textbf {\bibinfo {volume} {95}},\ \bibinfo {pages} {014906}
  (\bibinfo {year} {2017})},\ \Eprint {http://arxiv.org/abs/1609.02590}
  {arXiv:1609.02590 [nucl-th]} \BibitemShut {NoStop}%
\bibitem [{\citenamefont {Bozek}(2012)}]{Bozek:2011if}%
  \BibitemOpen
  \bibfield  {author} {\bibinfo {author} {\bibfnamefont {P.}~\bibnamefont
  {Bozek}},\ }\href {\doibase 10.1103/PhysRevC.85.014911} {\bibfield  {journal}
  {\bibinfo  {journal} {Phys. Rev. C}\ }\textbf {\bibinfo {volume} {85}},\
  \bibinfo {pages} {014911} (\bibinfo {year} {2012})},\ \Eprint
  {http://arxiv.org/abs/1112.0915} {arXiv:1112.0915 [hep-ph]} \BibitemShut
  {NoStop}%
\bibitem [{\citenamefont {Bzdak}\ \emph {et~al.}(2013)\citenamefont {Bzdak},
  \citenamefont {Schenke}, \citenamefont {Tribedy},\ and\ \citenamefont
  {Venugopalan}}]{Bzdak:2013zma}%
  \BibitemOpen
  \bibfield  {author} {\bibinfo {author} {\bibfnamefont {A.}~\bibnamefont
  {Bzdak}}, \bibinfo {author} {\bibfnamefont {B.}~\bibnamefont {Schenke}},
  \bibinfo {author} {\bibfnamefont {P.}~\bibnamefont {Tribedy}}, \ and\
  \bibinfo {author} {\bibfnamefont {R.}~\bibnamefont {Venugopalan}},\ }\href
  {\doibase 10.1103/PhysRevC.87.064906} {\bibfield  {journal} {\bibinfo
  {journal} {Phys. Rev. C}\ }\textbf {\bibinfo {volume} {87}},\ \bibinfo
  {pages} {064906} (\bibinfo {year} {2013})},\ \Eprint
  {http://arxiv.org/abs/1304.3403} {arXiv:1304.3403 [nucl-th]} \BibitemShut
  {NoStop}%
\bibitem [{\citenamefont {Qin}\ and\ \citenamefont
  {M{\"u}ller}(2014)}]{Qin:2013bha}%
  \BibitemOpen
  \bibfield  {author} {\bibinfo {author} {\bibfnamefont {G.-Y.}\ \bibnamefont
  {Qin}}\ and\ \bibinfo {author} {\bibfnamefont {B.}~\bibnamefont
  {M{\"u}ller}},\ }\href {\doibase 10.1103/PhysRevC.89.044902} {\bibfield
  {journal} {\bibinfo  {journal} {Phys. Rev. C}\ }\textbf {\bibinfo {volume}
  {89}},\ \bibinfo {pages} {044902} (\bibinfo {year} {2014})},\ \Eprint
  {http://arxiv.org/abs/1306.3439} {arXiv:1306.3439 [nucl-th]} \BibitemShut
  {NoStop}%
\bibitem [{\citenamefont {Nagle}\ \emph {et~al.}(2014)\citenamefont {Nagle},
  \citenamefont {Adare}, \citenamefont {Beckman}, \citenamefont {Koblesky},
  \citenamefont {Orjuela~Koop}, \citenamefont {McGlinchey}, \citenamefont
  {Romatschke}, \citenamefont {Carlson}, \citenamefont {Lynn},\ and\
  \citenamefont {McCumber}}]{Nagle:2013lja}%
  \BibitemOpen
  \bibfield  {author} {\bibinfo {author} {\bibfnamefont {J.~L.}\ \bibnamefont
  {Nagle}}, \bibinfo {author} {\bibfnamefont {A.}~\bibnamefont {Adare}},
  \bibinfo {author} {\bibfnamefont {S.}~\bibnamefont {Beckman}}, \bibinfo
  {author} {\bibfnamefont {T.}~\bibnamefont {Koblesky}}, \bibinfo {author}
  {\bibfnamefont {J.}~\bibnamefont {Orjuela~Koop}}, \bibinfo {author}
  {\bibfnamefont {D.}~\bibnamefont {McGlinchey}}, \bibinfo {author}
  {\bibfnamefont {P.}~\bibnamefont {Romatschke}}, \bibinfo {author}
  {\bibfnamefont {J.}~\bibnamefont {Carlson}}, \bibinfo {author} {\bibfnamefont
  {J.~E.}\ \bibnamefont {Lynn}}, \ and\ \bibinfo {author} {\bibfnamefont
  {M.}~\bibnamefont {McCumber}},\ }\href {\doibase
  10.1103/PhysRevLett.113.112301} {\bibfield  {journal} {\bibinfo  {journal}
  {Phys. Rev. Lett.}\ }\textbf {\bibinfo {volume} {113}},\ \bibinfo {pages}
  {112301} (\bibinfo {year} {2014})},\ \Eprint {http://arxiv.org/abs/1312.4565}
  {arXiv:1312.4565 [nucl-th]} \BibitemShut {NoStop}%
\bibitem [{\citenamefont {Werner}\ \emph
  {et~al.}(2014{\natexlab{a}})\citenamefont {Werner}, \citenamefont {Guiot},
  \citenamefont {Karpenko},\ and\ \citenamefont {Pierog}}]{Werner:2013tya}%
  \BibitemOpen
  \bibfield  {author} {\bibinfo {author} {\bibfnamefont {K.}~\bibnamefont
  {Werner}}, \bibinfo {author} {\bibfnamefont {B.}~\bibnamefont {Guiot}},
  \bibinfo {author} {\bibfnamefont {I.}~\bibnamefont {Karpenko}}, \ and\
  \bibinfo {author} {\bibfnamefont {T.}~\bibnamefont {Pierog}},\ }\href
  {\doibase 10.1103/PhysRevC.89.064903} {\bibfield  {journal} {\bibinfo
  {journal} {Phys. Rev. C}\ }\textbf {\bibinfo {volume} {89}},\ \bibinfo
  {pages} {064903} (\bibinfo {year} {2014}{\natexlab{a}})},\ \Eprint
  {http://arxiv.org/abs/1312.1233} {arXiv:1312.1233 [nucl-th]} \BibitemShut
  {NoStop}%
\bibitem [{\citenamefont {Werner}\ \emph
  {et~al.}(2014{\natexlab{b}})\citenamefont {Werner}, \citenamefont {Bleicher},
  \citenamefont {Guiot}, \citenamefont {Karpenko},\ and\ \citenamefont
  {Pierog}}]{Werner:2013ipa}%
  \BibitemOpen
  \bibfield  {author} {\bibinfo {author} {\bibfnamefont {K.}~\bibnamefont
  {Werner}}, \bibinfo {author} {\bibfnamefont {M.}~\bibnamefont {Bleicher}},
  \bibinfo {author} {\bibfnamefont {B.}~\bibnamefont {Guiot}}, \bibinfo
  {author} {\bibfnamefont {I.}~\bibnamefont {Karpenko}}, \ and\ \bibinfo
  {author} {\bibfnamefont {T.}~\bibnamefont {Pierog}},\ }\href {\doibase
  10.1103/PhysRevLett.112.232301} {\bibfield  {journal} {\bibinfo  {journal}
  {Phys. Rev. Lett.}\ }\textbf {\bibinfo {volume} {112}},\ \bibinfo {pages}
  {232301} (\bibinfo {year} {2014}{\natexlab{b}})},\ \Eprint
  {http://arxiv.org/abs/1307.4379} {arXiv:1307.4379 [nucl-th]} \BibitemShut
  {NoStop}%
\bibitem [{\citenamefont {Bozek}\ \emph {et~al.}(2013)\citenamefont {Bozek},
  \citenamefont {Broniowski},\ and\ \citenamefont {Torrieri}}]{Bozek:2013ska}%
  \BibitemOpen
  \bibfield  {author} {\bibinfo {author} {\bibfnamefont {P.}~\bibnamefont
  {Bozek}}, \bibinfo {author} {\bibfnamefont {W.}~\bibnamefont {Broniowski}}, \
  and\ \bibinfo {author} {\bibfnamefont {G.}~\bibnamefont {Torrieri}},\ }\href
  {\doibase 10.1103/PhysRevLett.111.172303} {\bibfield  {journal} {\bibinfo
  {journal} {Phys. Rev. Lett.}\ }\textbf {\bibinfo {volume} {111}},\ \bibinfo
  {pages} {172303} (\bibinfo {year} {2013})},\ \Eprint
  {http://arxiv.org/abs/1307.5060} {arXiv:1307.5060 [nucl-th]} \BibitemShut
  {NoStop}%
\bibitem [{\citenamefont {Schenke}\ and\ \citenamefont
  {Venugopalan}(2014)}]{Schenke:2014zha}%
  \BibitemOpen
  \bibfield  {author} {\bibinfo {author} {\bibfnamefont {B.}~\bibnamefont
  {Schenke}}\ and\ \bibinfo {author} {\bibfnamefont {R.}~\bibnamefont
  {Venugopalan}},\ }\href {\doibase 10.1103/PhysRevLett.113.102301} {\bibfield
  {journal} {\bibinfo  {journal} {Phys. Rev. Lett.}\ }\textbf {\bibinfo
  {volume} {113}},\ \bibinfo {pages} {102301} (\bibinfo {year} {2014})},\
  \Eprint {http://arxiv.org/abs/1405.3605} {arXiv:1405.3605 [nucl-th]}
  \BibitemShut {NoStop}%
\bibitem [{\citenamefont {Bozek}\ and\ \citenamefont
  {Broniowski}(2014)}]{Bozek:2014cya}%
  \BibitemOpen
  \bibfield  {author} {\bibinfo {author} {\bibfnamefont {P.}~\bibnamefont
  {Bozek}}\ and\ \bibinfo {author} {\bibfnamefont {W.}~\bibnamefont
  {Broniowski}},\ }\href {\doibase 10.1016/j.physletb.2014.11.006} {\bibfield
  {journal} {\bibinfo  {journal} {Phys. Lett. B}\ }\textbf {\bibinfo {volume}
  {739}},\ \bibinfo {pages} {308} (\bibinfo {year} {2014})},\ \Eprint
  {http://arxiv.org/abs/1409.2160} {arXiv:1409.2160 [nucl-th]} \BibitemShut
  {NoStop}%
\bibitem [{\citenamefont {Bozek}\ \emph {et~al.}(2015)\citenamefont {Bozek},
  \citenamefont {Bzdak},\ and\ \citenamefont {Ma}}]{Bozek:2015swa}%
  \BibitemOpen
  \bibfield  {author} {\bibinfo {author} {\bibfnamefont {P.}~\bibnamefont
  {Bozek}}, \bibinfo {author} {\bibfnamefont {A.}~\bibnamefont {Bzdak}}, \ and\
  \bibinfo {author} {\bibfnamefont {G.-L.}\ \bibnamefont {Ma}},\ }\href
  {\doibase 10.1016/j.physletb.2015.06.007} {\bibfield  {journal} {\bibinfo
  {journal} {Phys. Lett. B}\ }\textbf {\bibinfo {volume} {748}},\ \bibinfo
  {pages} {301} (\bibinfo {year} {2015})},\ \Eprint
  {http://arxiv.org/abs/1503.03655} {arXiv:1503.03655 [hep-ph]} \BibitemShut
  {NoStop}%
\bibitem [{\citenamefont {Zhou}\ \emph {et~al.}(2015)\citenamefont {Zhou},
  \citenamefont {Zhu}, \citenamefont {Li},\ and\ \citenamefont
  {Song}}]{Zhou:2015iba}%
  \BibitemOpen
  \bibfield  {author} {\bibinfo {author} {\bibfnamefont {Y.}~\bibnamefont
  {Zhou}}, \bibinfo {author} {\bibfnamefont {X.}~\bibnamefont {Zhu}}, \bibinfo
  {author} {\bibfnamefont {P.}~\bibnamefont {Li}}, \ and\ \bibinfo {author}
  {\bibfnamefont {H.}~\bibnamefont {Song}},\ }\href {\doibase
  10.1103/PhysRevC.91.064908} {\bibfield  {journal} {\bibinfo  {journal} {Phys.
  Rev. C}\ }\textbf {\bibinfo {volume} {91}},\ \bibinfo {pages} {064908}
  (\bibinfo {year} {2015})},\ \Eprint {http://arxiv.org/abs/1503.06986}
  {arXiv:1503.06986 [nucl-th]} \BibitemShut {NoStop}%
\bibitem [{\citenamefont {Weller}\ and\ \citenamefont
  {Romatschke}(2017)}]{Weller:2017tsr}%
  \BibitemOpen
  \bibfield  {author} {\bibinfo {author} {\bibfnamefont {R.~D.}\ \bibnamefont
  {Weller}}\ and\ \bibinfo {author} {\bibfnamefont {P.}~\bibnamefont
  {Romatschke}},\ }\href {\doibase 10.1016/j.physletb.2017.09.077} {\bibfield
  {journal} {\bibinfo  {journal} {Phys. Lett. B}\ }\textbf {\bibinfo {volume}
  {774}},\ \bibinfo {pages} {351} (\bibinfo {year} {2017})},\ \Eprint
  {http://arxiv.org/abs/1701.07145} {arXiv:1701.07145 [nucl-th]} \BibitemShut
  {NoStop}%
\bibitem [{\citenamefont {M{\"a}ntysaari}\ \emph {et~al.}(2017)\citenamefont
  {M{\"a}ntysaari}, \citenamefont {Schenke}, \citenamefont {Shen},\ and\
  \citenamefont {Tribedy}}]{Mantysaari:2017cni}%
  \BibitemOpen
  \bibfield  {author} {\bibinfo {author} {\bibfnamefont {H.}~\bibnamefont
  {M{\"a}ntysaari}}, \bibinfo {author} {\bibfnamefont {B.}~\bibnamefont
  {Schenke}}, \bibinfo {author} {\bibfnamefont {C.}~\bibnamefont {Shen}}, \
  and\ \bibinfo {author} {\bibfnamefont {P.}~\bibnamefont {Tribedy}},\ }\href
  {\doibase 10.1016/j.physletb.2017.07.038} {\bibfield  {journal} {\bibinfo
  {journal} {Phys. Lett. B}\ }\textbf {\bibinfo {volume} {772}},\ \bibinfo
  {pages} {681} (\bibinfo {year} {2017})},\ \Eprint
  {http://arxiv.org/abs/1705.03177} {arXiv:1705.03177 [nucl-th]} \BibitemShut
  {NoStop}%
\bibitem [{\citenamefont {Zhao}\ \emph {et~al.}(2018)\citenamefont {Zhao},
  \citenamefont {Zhou}, \citenamefont {Xu}, \citenamefont {Deng},\ and\
  \citenamefont {Song}}]{Zhao:2017rgg}%
  \BibitemOpen
  \bibfield  {author} {\bibinfo {author} {\bibfnamefont {W.}~\bibnamefont
  {Zhao}}, \bibinfo {author} {\bibfnamefont {Y.}~\bibnamefont {Zhou}}, \bibinfo
  {author} {\bibfnamefont {H.}~\bibnamefont {Xu}}, \bibinfo {author}
  {\bibfnamefont {W.}~\bibnamefont {Deng}}, \ and\ \bibinfo {author}
  {\bibfnamefont {H.}~\bibnamefont {Song}},\ }\href {\doibase
  10.1016/j.physletb.2018.03.022} {\bibfield  {journal} {\bibinfo  {journal}
  {Phys. Lett. B}\ }\textbf {\bibinfo {volume} {780}},\ \bibinfo {pages} {495}
  (\bibinfo {year} {2018})},\ \Eprint {http://arxiv.org/abs/1801.00271}
  {arXiv:1801.00271 [nucl-th]} \BibitemShut {NoStop}%
\bibitem [{\citenamefont {Schenke}\ \emph
  {et~al.}(2020{\natexlab{a}})\citenamefont {Schenke}, \citenamefont {Shen},\
  and\ \citenamefont {Tribedy}}]{Schenke:2020mbo}%
  \BibitemOpen
  \bibfield  {author} {\bibinfo {author} {\bibfnamefont {B.}~\bibnamefont
  {Schenke}}, \bibinfo {author} {\bibfnamefont {C.}~\bibnamefont {Shen}}, \
  and\ \bibinfo {author} {\bibfnamefont {P.}~\bibnamefont {Tribedy}},\ }\href
  {\doibase 10.1103/PhysRevC.102.044905} {\bibfield  {journal} {\bibinfo
  {journal} {Phys. Rev. C}\ }\textbf {\bibinfo {volume} {102}},\ \bibinfo
  {pages} {044905} (\bibinfo {year} {2020}{\natexlab{a}})},\ \Eprint
  {http://arxiv.org/abs/2005.14682} {arXiv:2005.14682 [nucl-th]} \BibitemShut
  {NoStop}%
\bibitem [{\citenamefont {Schenke}\ \emph
  {et~al.}(2020{\natexlab{b}})\citenamefont {Schenke}, \citenamefont {Shen},\
  and\ \citenamefont {Tribedy}}]{Schenke:2019pmk}%
  \BibitemOpen
  \bibfield  {author} {\bibinfo {author} {\bibfnamefont {B.}~\bibnamefont
  {Schenke}}, \bibinfo {author} {\bibfnamefont {C.}~\bibnamefont {Shen}}, \
  and\ \bibinfo {author} {\bibfnamefont {P.}~\bibnamefont {Tribedy}},\ }\href
  {\doibase 10.1016/j.physletb.2020.135322} {\bibfield  {journal} {\bibinfo
  {journal} {Phys. Lett. B}\ }\textbf {\bibinfo {volume} {803}},\ \bibinfo
  {pages} {135322} (\bibinfo {year} {2020}{\natexlab{b}})},\ \Eprint
  {http://arxiv.org/abs/1908.06212} {arXiv:1908.06212 [nucl-th]} \BibitemShut
  {NoStop}%
\bibitem [{\citenamefont {Orjuela~Koop}\ \emph {et~al.}(2016)\citenamefont
  {Orjuela~Koop}, \citenamefont {Belmont}, \citenamefont {Yin},\ and\
  \citenamefont {Nagle}}]{OrjuelaKoop:2015etn}%
  \BibitemOpen
  \bibfield  {author} {\bibinfo {author} {\bibfnamefont {J.~D.}\ \bibnamefont
  {Orjuela~Koop}}, \bibinfo {author} {\bibfnamefont {R.}~\bibnamefont
  {Belmont}}, \bibinfo {author} {\bibfnamefont {P.}~\bibnamefont {Yin}}, \ and\
  \bibinfo {author} {\bibfnamefont {J.~L.}\ \bibnamefont {Nagle}},\ }\href
  {\doibase 10.1103/PhysRevC.93.044910} {\bibfield  {journal} {\bibinfo
  {journal} {Phys. Rev. C}\ }\textbf {\bibinfo {volume} {93}},\ \bibinfo
  {pages} {044910} (\bibinfo {year} {2016})},\ \Eprint
  {http://arxiv.org/abs/1512.06949} {arXiv:1512.06949 [nucl-th]} \BibitemShut
  {NoStop}%
\bibitem [{\citenamefont {Bozek}\ and\ \citenamefont
  {Broniowski}(2015)}]{Bozek:2015qpa}%
  \BibitemOpen
  \bibfield  {author} {\bibinfo {author} {\bibfnamefont {P.}~\bibnamefont
  {Bozek}}\ and\ \bibinfo {author} {\bibfnamefont {W.}~\bibnamefont
  {Broniowski}},\ }\href {\doibase 10.1016/j.physletb.2015.05.068} {\bibfield
  {journal} {\bibinfo  {journal} {Phys. Lett. B}\ }\textbf {\bibinfo {volume}
  {747}},\ \bibinfo {pages} {135} (\bibinfo {year} {2015})},\ \Eprint
  {http://arxiv.org/abs/1503.00468} {arXiv:1503.00468 [nucl-th]} \BibitemShut
  {NoStop}%
\bibitem [{\citenamefont {Zhao}\ \emph
  {et~al.}(2020{\natexlab{a}})\citenamefont {Zhao}, \citenamefont {Zhou},
  \citenamefont {Murase},\ and\ \citenamefont {Song}}]{Zhao:2020pty}%
  \BibitemOpen
  \bibfield  {author} {\bibinfo {author} {\bibfnamefont {W.}~\bibnamefont
  {Zhao}}, \bibinfo {author} {\bibfnamefont {Y.}~\bibnamefont {Zhou}}, \bibinfo
  {author} {\bibfnamefont {K.}~\bibnamefont {Murase}}, \ and\ \bibinfo {author}
  {\bibfnamefont {H.}~\bibnamefont {Song}},\ }\href {\doibase
  10.1140/epjc/s10052-020-8376-x} {\bibfield  {journal} {\bibinfo  {journal}
  {Eur. Phys. J. C}\ }\textbf {\bibinfo {volume} {80}},\ \bibinfo {pages} {846}
  (\bibinfo {year} {2020}{\natexlab{a}})},\ \Eprint
  {http://arxiv.org/abs/2001.06742} {arXiv:2001.06742 [nucl-th]} \BibitemShut
  {NoStop}%
\bibitem [{\citenamefont {Zhao}\ \emph
  {et~al.}(2020{\natexlab{b}})\citenamefont {Zhao}, \citenamefont {Ko},
  \citenamefont {Liu}, \citenamefont {Qin},\ and\ \citenamefont
  {Song}}]{Zhao:2020wcd}%
  \BibitemOpen
  \bibfield  {author} {\bibinfo {author} {\bibfnamefont {W.}~\bibnamefont
  {Zhao}}, \bibinfo {author} {\bibfnamefont {C.~M.}\ \bibnamefont {Ko}},
  \bibinfo {author} {\bibfnamefont {Y.-X.}\ \bibnamefont {Liu}}, \bibinfo
  {author} {\bibfnamefont {G.-Y.}\ \bibnamefont {Qin}}, \ and\ \bibinfo
  {author} {\bibfnamefont {H.}~\bibnamefont {Song}},\ }\href {\doibase
  10.1103/PhysRevLett.125.072301} {\bibfield  {journal} {\bibinfo  {journal}
  {Phys. Rev. Lett.}\ }\textbf {\bibinfo {volume} {125}},\ \bibinfo {pages}
  {072301} (\bibinfo {year} {2020}{\natexlab{b}})},\ \Eprint
  {http://arxiv.org/abs/1911.00826} {arXiv:1911.00826 [nucl-th]} \BibitemShut
  {NoStop}%
\bibitem [{\citenamefont {Zhao}\ \emph {et~al.}(2022)\citenamefont {Zhao},
  \citenamefont {Shen},\ and\ \citenamefont {Schenke}}]{Zhao:2022ayk}%
  \BibitemOpen
  \bibfield  {author} {\bibinfo {author} {\bibfnamefont {W.}~\bibnamefont
  {Zhao}}, \bibinfo {author} {\bibfnamefont {C.}~\bibnamefont {Shen}}, \ and\
  \bibinfo {author} {\bibfnamefont {B.}~\bibnamefont {Schenke}},\ }\href
  {\doibase 10.1103/PhysRevLett.129.252302} {\bibfield  {journal} {\bibinfo
  {journal} {Phys. Rev. Lett.}\ }\textbf {\bibinfo {volume} {129}},\ \bibinfo
  {pages} {252302} (\bibinfo {year} {2022})},\ \Eprint
  {http://arxiv.org/abs/2203.06094} {arXiv:2203.06094 [nucl-th]} \BibitemShut
  {NoStop}%
\bibitem [{\citenamefont {Denicol}\ \emph {et~al.}(2014)\citenamefont
  {Denicol}, \citenamefont {Heinz}, \citenamefont {Martinez}, \citenamefont
  {Noronha},\ and\ \citenamefont {Strickland}}]{Denicol:2014tha}%
  \BibitemOpen
  \bibfield  {author} {\bibinfo {author} {\bibfnamefont {G.~S.}\ \bibnamefont
  {Denicol}}, \bibinfo {author} {\bibfnamefont {U.}~\bibnamefont {Heinz}},
  \bibinfo {author} {\bibfnamefont {M.}~\bibnamefont {Martinez}}, \bibinfo
  {author} {\bibfnamefont {J.}~\bibnamefont {Noronha}}, \ and\ \bibinfo
  {author} {\bibfnamefont {M.}~\bibnamefont {Strickland}},\ }\href {\doibase
  10.1103/PhysRevD.90.125026} {\bibfield  {journal} {\bibinfo  {journal} {Phys.
  Rev. D}\ }\textbf {\bibinfo {volume} {90}},\ \bibinfo {pages} {125026}
  (\bibinfo {year} {2014})},\ \Eprint {http://arxiv.org/abs/1408.7048}
  {arXiv:1408.7048 [hep-ph]} \BibitemShut {NoStop}%
\bibitem [{\citenamefont {Kurkela}\ \emph
  {et~al.}(2019{\natexlab{a}})\citenamefont {Kurkela}, \citenamefont
  {Mazeliauskas}, \citenamefont {Paquet}, \citenamefont {Schlichting},\ and\
  \citenamefont {Teaney}}]{Kurkela:2018wud}%
  \BibitemOpen
  \bibfield  {author} {\bibinfo {author} {\bibfnamefont {A.}~\bibnamefont
  {Kurkela}}, \bibinfo {author} {\bibfnamefont {A.}~\bibnamefont
  {Mazeliauskas}}, \bibinfo {author} {\bibfnamefont {J.-F.}\ \bibnamefont
  {Paquet}}, \bibinfo {author} {\bibfnamefont {S.}~\bibnamefont {Schlichting}},
  \ and\ \bibinfo {author} {\bibfnamefont {D.}~\bibnamefont {Teaney}},\ }\href
  {\doibase 10.1103/PhysRevLett.122.122302} {\bibfield  {journal} {\bibinfo
  {journal} {Phys. Rev. Lett.}\ }\textbf {\bibinfo {volume} {122}},\ \bibinfo
  {pages} {122302} (\bibinfo {year} {2019}{\natexlab{a}})},\ \Eprint
  {http://arxiv.org/abs/1805.01604} {arXiv:1805.01604 [hep-ph]} \BibitemShut
  {NoStop}%
\bibitem [{\citenamefont {Kurkela}\ and\ \citenamefont
  {Mazeliauskas}(2020)}]{Kurkela:2019cgr}%
  \BibitemOpen
  \bibfield  {author} {\bibinfo {author} {\bibfnamefont {A.}~\bibnamefont
  {Kurkela}}\ and\ \bibinfo {author} {\bibfnamefont {A.}~\bibnamefont
  {Mazeliauskas}},\ }\href {\doibase 10.1007/978-3-030-53448-6_26} {\bibfield
  {journal} {\bibinfo  {journal} {Springer Proc. Phys.}\ }\textbf {\bibinfo
  {volume} {250}},\ \bibinfo {pages} {177} (\bibinfo {year} {2020})},\ \Eprint
  {http://arxiv.org/abs/1910.06664} {arXiv:1910.06664 [hep-ph]} \BibitemShut
  {NoStop}%
\bibitem [{\citenamefont {Kurkela}\ \emph
  {et~al.}(2019{\natexlab{b}})\citenamefont {Kurkela}, \citenamefont
  {Wiedemann},\ and\ \citenamefont {Wu}}]{Kurkela:2019kip}%
  \BibitemOpen
  \bibfield  {author} {\bibinfo {author} {\bibfnamefont {A.}~\bibnamefont
  {Kurkela}}, \bibinfo {author} {\bibfnamefont {U.~A.}\ \bibnamefont
  {Wiedemann}}, \ and\ \bibinfo {author} {\bibfnamefont {B.}~\bibnamefont
  {Wu}},\ }\href {\doibase 10.1140/epjc/s10052-019-7428-6} {\bibfield
  {journal} {\bibinfo  {journal} {Eur. Phys. J. C}\ }\textbf {\bibinfo {volume}
  {79}},\ \bibinfo {pages} {965} (\bibinfo {year} {2019}{\natexlab{b}})},\
  \Eprint {http://arxiv.org/abs/1905.05139} {arXiv:1905.05139 [hep-ph]}
  \BibitemShut {NoStop}%
\bibitem [{\citenamefont {Kurkela}\ \emph {et~al.}(2020)\citenamefont
  {Kurkela}, \citenamefont {Taghavi}, \citenamefont {Wiedemann},\ and\
  \citenamefont {Wu}}]{Kurkela:2020wwb}%
  \BibitemOpen
  \bibfield  {author} {\bibinfo {author} {\bibfnamefont {A.}~\bibnamefont
  {Kurkela}}, \bibinfo {author} {\bibfnamefont {S.~F.}\ \bibnamefont
  {Taghavi}}, \bibinfo {author} {\bibfnamefont {U.~A.}\ \bibnamefont
  {Wiedemann}}, \ and\ \bibinfo {author} {\bibfnamefont {B.}~\bibnamefont
  {Wu}},\ }\href {\doibase 10.1016/j.physletb.2020.135901} {\bibfield
  {journal} {\bibinfo  {journal} {Phys. Lett. B}\ }\textbf {\bibinfo {volume}
  {811}},\ \bibinfo {pages} {135901} (\bibinfo {year} {2020})},\ \Eprint
  {http://arxiv.org/abs/2007.06851} {arXiv:2007.06851 [hep-ph]} \BibitemShut
  {NoStop}%
\bibitem [{\citenamefont {Denicol}\ and\ \citenamefont
  {Noronha}(2020)}]{Denicol:2019lio}%
  \BibitemOpen
  \bibfield  {author} {\bibinfo {author} {\bibfnamefont {G.~S.}\ \bibnamefont
  {Denicol}}\ and\ \bibinfo {author} {\bibfnamefont {J.}~\bibnamefont
  {Noronha}},\ }\href {\doibase 10.1103/PhysRevLett.124.152301} {\bibfield
  {journal} {\bibinfo  {journal} {Phys. Rev. Lett.}\ }\textbf {\bibinfo
  {volume} {124}},\ \bibinfo {pages} {152301} (\bibinfo {year} {2020})},\
  \Eprint {http://arxiv.org/abs/1908.09957} {arXiv:1908.09957 [nucl-th]}
  \BibitemShut {NoStop}%
\bibitem [{\citenamefont {Du}\ and\ \citenamefont
  {Schlichting}(2021)}]{Du:2020zqg}%
  \BibitemOpen
  \bibfield  {author} {\bibinfo {author} {\bibfnamefont {X.}~\bibnamefont
  {Du}}\ and\ \bibinfo {author} {\bibfnamefont {S.}~\bibnamefont
  {Schlichting}},\ }\href {\doibase 10.1103/PhysRevLett.127.122301} {\bibfield
  {journal} {\bibinfo  {journal} {Phys. Rev. Lett.}\ }\textbf {\bibinfo
  {volume} {127}},\ \bibinfo {pages} {122301} (\bibinfo {year} {2021})},\
  \Eprint {http://arxiv.org/abs/2012.09068} {arXiv:2012.09068 [hep-ph]}
  \BibitemShut {NoStop}%
\bibitem [{\citenamefont {Du}\ \emph {et~al.}(2022)\citenamefont {Du},
  \citenamefont {Heller}, \citenamefont {Schlichting},\ and\ \citenamefont
  {Svensson}}]{Du:2022bel}%
  \BibitemOpen
  \bibfield  {author} {\bibinfo {author} {\bibfnamefont {X.}~\bibnamefont
  {Du}}, \bibinfo {author} {\bibfnamefont {M.~P.}\ \bibnamefont {Heller}},
  \bibinfo {author} {\bibfnamefont {S.}~\bibnamefont {Schlichting}}, \ and\
  \bibinfo {author} {\bibfnamefont {V.}~\bibnamefont {Svensson}},\ }\href
  {\doibase 10.1103/PhysRevD.106.014016} {\bibfield  {journal} {\bibinfo
  {journal} {Phys. Rev. D}\ }\textbf {\bibinfo {volume} {106}},\ \bibinfo
  {pages} {014016} (\bibinfo {year} {2022})},\ \Eprint
  {http://arxiv.org/abs/2203.16549} {arXiv:2203.16549 [hep-ph]} \BibitemShut
  {NoStop}%
\bibitem [{\citenamefont {Blaizot}\ and\ \citenamefont
  {Yan}(2021)}]{Blaizot:2021cdv}%
  \BibitemOpen
  \bibfield  {author} {\bibinfo {author} {\bibfnamefont {J.-P.}\ \bibnamefont
  {Blaizot}}\ and\ \bibinfo {author} {\bibfnamefont {L.}~\bibnamefont {Yan}},\
  }\href {\doibase 10.1103/PhysRevC.104.055201} {\bibfield  {journal} {\bibinfo
   {journal} {Phys. Rev. C}\ }\textbf {\bibinfo {volume} {104}},\ \bibinfo
  {pages} {055201} (\bibinfo {year} {2021})},\ \Eprint
  {http://arxiv.org/abs/2106.10508} {arXiv:2106.10508 [nucl-th]} \BibitemShut
  {NoStop}%
\bibitem [{\citenamefont {Ambrus}\ \emph
  {et~al.}(2023{\natexlab{a}})\citenamefont {Ambrus}, \citenamefont
  {Schlichting},\ and\ \citenamefont {Werthmann}}]{Ambrus:2022koq}%
  \BibitemOpen
  \bibfield  {author} {\bibinfo {author} {\bibfnamefont {V.~E.}\ \bibnamefont
  {Ambrus}}, \bibinfo {author} {\bibfnamefont {S.}~\bibnamefont {Schlichting}},
  \ and\ \bibinfo {author} {\bibfnamefont {C.}~\bibnamefont {Werthmann}},\
  }\href {\doibase 10.1103/PhysRevD.107.094013} {\bibfield  {journal} {\bibinfo
   {journal} {Phys. Rev. D}\ }\textbf {\bibinfo {volume} {107}},\ \bibinfo
  {pages} {094013} (\bibinfo {year} {2023}{\natexlab{a}})},\ \Eprint
  {http://arxiv.org/abs/2211.14379} {arXiv:2211.14379 [hep-ph]} \BibitemShut
  {NoStop}%
\bibitem [{\citenamefont {Arslandok}\ \emph {et~al.}(2023)\citenamefont
  {Arslandok} \emph {et~al.}}]{Arslandok:2023utm}%
  \BibitemOpen
  \bibfield  {author} {\bibinfo {author} {\bibfnamefont {M.}~\bibnamefont
  {Arslandok}} \emph {et~al.},\ }\href@noop {} {\  (\bibinfo {year} {2023})},\
  \Eprint {http://arxiv.org/abs/2303.17254} {arXiv:2303.17254 [nucl-ex]}
  \BibitemShut {NoStop}%
\bibitem [{\citenamefont {Ambrus}\ \emph
  {et~al.}(2023{\natexlab{b}})\citenamefont {Ambrus}, \citenamefont
  {Schlichting},\ and\ \citenamefont {Werthmann}}]{Ambrus:2022qya}%
  \BibitemOpen
  \bibfield  {author} {\bibinfo {author} {\bibfnamefont {V.~E.}\ \bibnamefont
  {Ambrus}}, \bibinfo {author} {\bibfnamefont {S.}~\bibnamefont {Schlichting}},
  \ and\ \bibinfo {author} {\bibfnamefont {C.}~\bibnamefont {Werthmann}},\
  }\href {\doibase 10.1103/PhysRevLett.130.152301} {\bibfield  {journal}
  {\bibinfo  {journal} {Phys. Rev. Lett.}\ }\textbf {\bibinfo {volume} {130}},\
  \bibinfo {pages} {152301} (\bibinfo {year} {2023}{\natexlab{b}})},\ \Eprint
  {http://arxiv.org/abs/2211.14356} {arXiv:2211.14356 [hep-ph]} \BibitemShut
  {NoStop}%
\bibitem [{\citenamefont {Werthmann}\ \emph {et~al.}(2024)\citenamefont
  {Werthmann}, \citenamefont {Ambrus},\ and\ \citenamefont
  {Schlichting}}]{Werthmann:2023dvl}%
  \BibitemOpen
  \bibfield  {author} {\bibinfo {author} {\bibfnamefont {C.}~\bibnamefont
  {Werthmann}}, \bibinfo {author} {\bibfnamefont {V.~E.}\ \bibnamefont
  {Ambrus}}, \ and\ \bibinfo {author} {\bibfnamefont {S.}~\bibnamefont
  {Schlichting}},\ }\href {\doibase 10.22323/1.438.0048} {\bibfield  {journal}
  {\bibinfo  {journal} {PoS}\ }\textbf {\bibinfo {volume} {HardProbes2023}},\
  \bibinfo {pages} {048} (\bibinfo {year} {2024})},\ \Eprint
  {http://arxiv.org/abs/2307.08306} {arXiv:2307.08306 [hep-ph]} \BibitemShut
  {NoStop}%
\bibitem [{\citenamefont {Ambru\c{s}}\ \emph {et~al.}(2024)\citenamefont
  {Ambru\c{s}}, \citenamefont {Schlichting},\ and\ \citenamefont
  {Werthmann}}]{Ambrus:2023oyk}%
  \BibitemOpen
  \bibfield  {author} {\bibinfo {author} {\bibfnamefont {V.~E.}\ \bibnamefont
  {Ambru\c{s}}}, \bibinfo {author} {\bibfnamefont {S.}~\bibnamefont
  {Schlichting}}, \ and\ \bibinfo {author} {\bibfnamefont {C.}~\bibnamefont
  {Werthmann}},\ }\href {\doibase 10.1063/5.0215368} {\bibfield  {journal}
  {\bibinfo  {journal} {AIP Conf. Proc.}\ }\textbf {\bibinfo {volume} {3181}},\
  \bibinfo {pages} {050004} (\bibinfo {year} {2024})},\ \Eprint
  {http://arxiv.org/abs/2302.10618} {arXiv:2302.10618 [nucl-th]} \BibitemShut
  {NoStop}%
\bibitem [{\citenamefont {Ambrus}\ \emph
  {et~al.}(2025{\natexlab{a}})\citenamefont {Ambrus}, \citenamefont
  {Schlichting},\ and\ \citenamefont {Werthmann}}]{Ambrus:2024eqa}%
  \BibitemOpen
  \bibfield  {author} {\bibinfo {author} {\bibfnamefont {V.~E.}\ \bibnamefont
  {Ambrus}}, \bibinfo {author} {\bibfnamefont {S.}~\bibnamefont {Schlichting}},
  \ and\ \bibinfo {author} {\bibfnamefont {C.}~\bibnamefont {Werthmann}},\
  }\href {\doibase 10.1103/PhysRevD.111.054025} {\bibfield  {journal} {\bibinfo
   {journal} {Phys. Rev. D}\ }\textbf {\bibinfo {volume} {111}},\ \bibinfo
  {pages} {054025} (\bibinfo {year} {2025}{\natexlab{a}})},\ \Eprint
  {http://arxiv.org/abs/2411.19709} {arXiv:2411.19709 [hep-ph]} \BibitemShut
  {NoStop}%
\bibitem [{\citenamefont {Ambrus}\ \emph
  {et~al.}(2025{\natexlab{b}})\citenamefont {Ambrus}, \citenamefont
  {Schlichting},\ and\ \citenamefont {Werthmann}}]{Ambrus:2024hks}%
  \BibitemOpen
  \bibfield  {author} {\bibinfo {author} {\bibfnamefont {V.~E.}\ \bibnamefont
  {Ambrus}}, \bibinfo {author} {\bibfnamefont {S.}~\bibnamefont {Schlichting}},
  \ and\ \bibinfo {author} {\bibfnamefont {C.}~\bibnamefont {Werthmann}},\
  }\href {\doibase 10.1103/PhysRevD.111.054024} {\bibfield  {journal} {\bibinfo
   {journal} {Phys. Rev. D}\ }\textbf {\bibinfo {volume} {111}},\ \bibinfo
  {pages} {054024} (\bibinfo {year} {2025}{\natexlab{b}})},\ \Eprint
  {http://arxiv.org/abs/2411.19708} {arXiv:2411.19708 [hep-ph]} \BibitemShut
  {NoStop}%
\bibitem [{\citenamefont {Martinez}\ and\ \citenamefont
  {Strickland}(2010)}]{Martinez:2010sc}%
  \BibitemOpen
  \bibfield  {author} {\bibinfo {author} {\bibfnamefont {M.}~\bibnamefont
  {Martinez}}\ and\ \bibinfo {author} {\bibfnamefont {M.}~\bibnamefont
  {Strickland}},\ }\href {\doibase 10.1016/j.nuclphysa.2010.08.011} {\bibfield
  {journal} {\bibinfo  {journal} {Nucl. Phys. A}\ }\textbf {\bibinfo {volume}
  {848}},\ \bibinfo {pages} {183} (\bibinfo {year} {2010})},\ \Eprint
  {http://arxiv.org/abs/1007.0889} {arXiv:1007.0889 [nucl-th]} \BibitemShut
  {NoStop}%
\bibitem [{\citenamefont {Florkowski}\ \emph
  {et~al.}(2013{\natexlab{a}})\citenamefont {Florkowski}, \citenamefont
  {Martinez}, \citenamefont {Ryblewski},\ and\ \citenamefont
  {Strickland}}]{Florkowski:2012pf}%
  \BibitemOpen
  \bibfield  {author} {\bibinfo {author} {\bibfnamefont {W.}~\bibnamefont
  {Florkowski}}, \bibinfo {author} {\bibfnamefont {M.}~\bibnamefont
  {Martinez}}, \bibinfo {author} {\bibfnamefont {R.}~\bibnamefont {Ryblewski}},
  \ and\ \bibinfo {author} {\bibfnamefont {M.}~\bibnamefont {Strickland}},\
  }\href {\doibase 10.1016/j.nuclphysa.2013.02.138} {\bibfield  {journal}
  {\bibinfo  {journal} {Nucl. Phys. A}\ }\textbf {\bibinfo {volume}
  {904-905}},\ \bibinfo {pages} {803c} (\bibinfo {year}
  {2013}{\natexlab{a}})},\ \Eprint {http://arxiv.org/abs/1210.1677}
  {arXiv:1210.1677 [nucl-th]} \BibitemShut {NoStop}%
\bibitem [{\citenamefont {Florkowski}\ and\ \citenamefont
  {Madetko}(2014)}]{Florkowski:2014txa}%
  \BibitemOpen
  \bibfield  {author} {\bibinfo {author} {\bibfnamefont {W.}~\bibnamefont
  {Florkowski}}\ and\ \bibinfo {author} {\bibfnamefont {O.}~\bibnamefont
  {Madetko}},\ }\href {\doibase 10.5506/APhysPolB.45.1103} {\bibfield
  {journal} {\bibinfo  {journal} {Acta Phys. Polon. B}\ }\textbf {\bibinfo
  {volume} {45}},\ \bibinfo {pages} {1103} (\bibinfo {year} {2014})},\ \Eprint
  {http://arxiv.org/abs/1402.2401} {arXiv:1402.2401 [nucl-th]} \BibitemShut
  {NoStop}%
\bibitem [{\citenamefont {Strickland}(2014)}]{Strickland:2014pga}%
  \BibitemOpen
  \bibfield  {author} {\bibinfo {author} {\bibfnamefont {M.}~\bibnamefont
  {Strickland}},\ }\href {\doibase 10.5506/APhysPolB.45.2355} {\bibfield
  {journal} {\bibinfo  {journal} {Acta Phys. Polon. B}\ }\textbf {\bibinfo
  {volume} {45}},\ \bibinfo {pages} {2355} (\bibinfo {year} {2014})},\ \Eprint
  {http://arxiv.org/abs/1410.5786} {arXiv:1410.5786 [nucl-th]} \BibitemShut
  {NoStop}%
\bibitem [{\citenamefont {Kasmaei}\ and\ \citenamefont
  {Strickland}(2020)}]{Kasmaei:2019ofu}%
  \BibitemOpen
  \bibfield  {author} {\bibinfo {author} {\bibfnamefont {B.~S.}\ \bibnamefont
  {Kasmaei}}\ and\ \bibinfo {author} {\bibfnamefont {M.}~\bibnamefont
  {Strickland}},\ }\href {\doibase 10.1103/PhysRevD.102.014037} {\bibfield
  {journal} {\bibinfo  {journal} {Phys. Rev. D}\ }\textbf {\bibinfo {volume}
  {102}},\ \bibinfo {pages} {014037} (\bibinfo {year} {2020})},\ \Eprint
  {http://arxiv.org/abs/1911.03370} {arXiv:1911.03370 [hep-ph]} \BibitemShut
  {NoStop}%
\bibitem [{\citenamefont {Behtash}\ \emph {et~al.}(2018)\citenamefont
  {Behtash}, \citenamefont {Cruz-Camacho},\ and\ \citenamefont
  {Martinez}}]{Behtash:2017wqg}%
  \BibitemOpen
  \bibfield  {author} {\bibinfo {author} {\bibfnamefont {A.}~\bibnamefont
  {Behtash}}, \bibinfo {author} {\bibfnamefont {C.~N.}\ \bibnamefont
  {Cruz-Camacho}}, \ and\ \bibinfo {author} {\bibfnamefont {M.}~\bibnamefont
  {Martinez}},\ }\href {\doibase 10.1103/PhysRevD.97.044041} {\bibfield
  {journal} {\bibinfo  {journal} {Phys. Rev. D}\ }\textbf {\bibinfo {volume}
  {97}},\ \bibinfo {pages} {044041} (\bibinfo {year} {2018})},\ \Eprint
  {http://arxiv.org/abs/1711.01745} {arXiv:1711.01745 [hep-th]} \BibitemShut
  {NoStop}%
\bibitem [{\citenamefont {Strickland}(2024)}]{Strickland:2024moq}%
  \BibitemOpen
  \bibfield  {author} {\bibinfo {author} {\bibfnamefont {M.}~\bibnamefont
  {Strickland}},\ }\href {\doibase 10.1142/9789811294679_0003} {\bibfield
  {journal} {\bibinfo  {journal} {Int. J. Mod. Phys. E}\ }\textbf {\bibinfo
  {volume} {33}},\ \bibinfo {pages} {2430004} (\bibinfo {year} {2024})},\
  \Eprint {http://arxiv.org/abs/2402.09571} {arXiv:2402.09571 [nucl-th]}
  \BibitemShut {NoStop}%
\bibitem [{\citenamefont {Martinez}\ \emph {et~al.}(2012)\citenamefont
  {Martinez}, \citenamefont {Ryblewski},\ and\ \citenamefont
  {Strickland}}]{Martinez:2012tu}%
  \BibitemOpen
  \bibfield  {author} {\bibinfo {author} {\bibfnamefont {M.}~\bibnamefont
  {Martinez}}, \bibinfo {author} {\bibfnamefont {R.}~\bibnamefont {Ryblewski}},
  \ and\ \bibinfo {author} {\bibfnamefont {M.}~\bibnamefont {Strickland}},\
  }\href {\doibase 10.1103/PhysRevC.85.064913} {\bibfield  {journal} {\bibinfo
  {journal} {Phys. Rev. C}\ }\textbf {\bibinfo {volume} {85}},\ \bibinfo
  {pages} {064913} (\bibinfo {year} {2012})},\ \Eprint
  {http://arxiv.org/abs/1204.1473} {arXiv:1204.1473 [nucl-th]} \BibitemShut
  {NoStop}%
\bibitem [{\citenamefont {Tinti}(2015)}]{Tinti:2014yya}%
  \BibitemOpen
  \bibfield  {author} {\bibinfo {author} {\bibfnamefont {L.}~\bibnamefont
  {Tinti}},\ }\href {\doibase 10.1103/PhysRevC.92.014908} {\bibfield  {journal}
  {\bibinfo  {journal} {Phys. Rev. C}\ }\textbf {\bibinfo {volume} {92}},\
  \bibinfo {pages} {014908} (\bibinfo {year} {2015})},\ \Eprint
  {http://arxiv.org/abs/1411.7268} {arXiv:1411.7268 [nucl-th]} \BibitemShut
  {NoStop}%
\bibitem [{\citenamefont {Alqahtani}\ \emph {et~al.}(2017)\citenamefont
  {Alqahtani}, \citenamefont {Nopoush},\ and\ \citenamefont
  {Strickland}}]{Alqahtani:2016rth}%
  \BibitemOpen
  \bibfield  {author} {\bibinfo {author} {\bibfnamefont {M.}~\bibnamefont
  {Alqahtani}}, \bibinfo {author} {\bibfnamefont {M.}~\bibnamefont {Nopoush}},
  \ and\ \bibinfo {author} {\bibfnamefont {M.}~\bibnamefont {Strickland}},\
  }\href {\doibase 10.1103/PhysRevC.95.034906} {\bibfield  {journal} {\bibinfo
  {journal} {Phys. Rev. C}\ }\textbf {\bibinfo {volume} {95}},\ \bibinfo
  {pages} {034906} (\bibinfo {year} {2017})},\ \Eprint
  {http://arxiv.org/abs/1605.02101} {arXiv:1605.02101 [nucl-th]} \BibitemShut
  {NoStop}%
\bibitem [{\citenamefont {Bazow}\ \emph {et~al.}(2014)\citenamefont {Bazow},
  \citenamefont {Heinz},\ and\ \citenamefont {Strickland}}]{Bazow:2013ifa}%
  \BibitemOpen
  \bibfield  {author} {\bibinfo {author} {\bibfnamefont {D.}~\bibnamefont
  {Bazow}}, \bibinfo {author} {\bibfnamefont {U.}~\bibnamefont {Heinz}}, \ and\
  \bibinfo {author} {\bibfnamefont {M.}~\bibnamefont {Strickland}},\ }\href
  {\doibase 10.1103/PhysRevC.90.054910} {\bibfield  {journal} {\bibinfo
  {journal} {Phys. Rev. C}\ }\textbf {\bibinfo {volume} {90}},\ \bibinfo
  {pages} {054910} (\bibinfo {year} {2014})},\ \Eprint
  {http://arxiv.org/abs/1311.6720} {arXiv:1311.6720 [nucl-th]} \BibitemShut
  {NoStop}%
\bibitem [{\citenamefont {Molnar}\ \emph {et~al.}(2016)\citenamefont {Molnar},
  \citenamefont {Niemi},\ and\ \citenamefont {Rischke}}]{Molnar:2016vvu}%
  \BibitemOpen
  \bibfield  {author} {\bibinfo {author} {\bibfnamefont {E.}~\bibnamefont
  {Molnar}}, \bibinfo {author} {\bibfnamefont {H.}~\bibnamefont {Niemi}}, \
  and\ \bibinfo {author} {\bibfnamefont {D.~H.}\ \bibnamefont {Rischke}},\
  }\href {\doibase 10.1103/PhysRevD.93.114025} {\bibfield  {journal} {\bibinfo
  {journal} {Phys. Rev. D}\ }\textbf {\bibinfo {volume} {93}},\ \bibinfo
  {pages} {114025} (\bibinfo {year} {2016})},\ \Eprint
  {http://arxiv.org/abs/1602.00573} {arXiv:1602.00573 [nucl-th]} \BibitemShut
  {NoStop}%
\bibitem [{\citenamefont {McNelis}\ \emph {et~al.}(2018)\citenamefont
  {McNelis}, \citenamefont {Bazow},\ and\ \citenamefont
  {Heinz}}]{McNelis:2018jho}%
  \BibitemOpen
  \bibfield  {author} {\bibinfo {author} {\bibfnamefont {M.}~\bibnamefont
  {McNelis}}, \bibinfo {author} {\bibfnamefont {D.}~\bibnamefont {Bazow}}, \
  and\ \bibinfo {author} {\bibfnamefont {U.}~\bibnamefont {Heinz}},\ }\href
  {\doibase 10.1103/PhysRevC.97.054912} {\bibfield  {journal} {\bibinfo
  {journal} {Phys. Rev. C}\ }\textbf {\bibinfo {volume} {97}},\ \bibinfo
  {pages} {054912} (\bibinfo {year} {2018})},\ \Eprint
  {http://arxiv.org/abs/1803.01810} {arXiv:1803.01810 [nucl-th]} \BibitemShut
  {NoStop}%
\bibitem [{\citenamefont {Florkowski}\ and\ \citenamefont
  {Ryblewski}(2011)}]{Florkowski:2010cf}%
  \BibitemOpen
  \bibfield  {author} {\bibinfo {author} {\bibfnamefont {W.}~\bibnamefont
  {Florkowski}}\ and\ \bibinfo {author} {\bibfnamefont {R.}~\bibnamefont
  {Ryblewski}},\ }\href {\doibase 10.1103/PhysRevC.83.034907} {\bibfield
  {journal} {\bibinfo  {journal} {Phys. Rev. C}\ }\textbf {\bibinfo {volume}
  {83}},\ \bibinfo {pages} {034907} (\bibinfo {year} {2011})},\ \Eprint
  {http://arxiv.org/abs/1007.0130} {arXiv:1007.0130 [nucl-th]} \BibitemShut
  {NoStop}%
\bibitem [{\citenamefont {Florkowski}\ \emph {et~al.}(2015)\citenamefont
  {Florkowski}, \citenamefont {Maksymiuk}, \citenamefont {Ryblewski},\ and\
  \citenamefont {Tinti}}]{Florkowski:2015cba}%
  \BibitemOpen
  \bibfield  {author} {\bibinfo {author} {\bibfnamefont {W.}~\bibnamefont
  {Florkowski}}, \bibinfo {author} {\bibfnamefont {E.}~\bibnamefont
  {Maksymiuk}}, \bibinfo {author} {\bibfnamefont {R.}~\bibnamefont
  {Ryblewski}}, \ and\ \bibinfo {author} {\bibfnamefont {L.}~\bibnamefont
  {Tinti}},\ }\href {\doibase 10.1103/PhysRevC.92.054912} {\bibfield  {journal}
  {\bibinfo  {journal} {Phys. Rev. C}\ }\textbf {\bibinfo {volume} {92}},\
  \bibinfo {pages} {054912} (\bibinfo {year} {2015})},\ \Eprint
  {http://arxiv.org/abs/1508.04534} {arXiv:1508.04534 [nucl-th]} \BibitemShut
  {NoStop}%
\bibitem [{\citenamefont {Bazow}\ \emph {et~al.}(2015)\citenamefont {Bazow},
  \citenamefont {Heinz},\ and\ \citenamefont {Martinez}}]{Bazow:2015cha}%
  \BibitemOpen
  \bibfield  {author} {\bibinfo {author} {\bibfnamefont {D.}~\bibnamefont
  {Bazow}}, \bibinfo {author} {\bibfnamefont {U.}~\bibnamefont {Heinz}}, \ and\
  \bibinfo {author} {\bibfnamefont {M.}~\bibnamefont {Martinez}},\ }\href
  {\doibase 10.1103/PhysRevC.91.064903} {\bibfield  {journal} {\bibinfo
  {journal} {Phys. Rev. C}\ }\textbf {\bibinfo {volume} {91}},\ \bibinfo
  {pages} {064903} (\bibinfo {year} {2015})},\ \Eprint
  {http://arxiv.org/abs/1503.07443} {arXiv:1503.07443 [nucl-th]} \BibitemShut
  {NoStop}%
\bibitem [{\citenamefont {Tinti}(2016)}]{Tinti:2015xwa}%
  \BibitemOpen
  \bibfield  {author} {\bibinfo {author} {\bibfnamefont {L.}~\bibnamefont
  {Tinti}},\ }\href {\doibase 10.1103/PhysRevC.94.044902} {\bibfield  {journal}
  {\bibinfo  {journal} {Phys. Rev. C}\ }\textbf {\bibinfo {volume} {94}},\
  \bibinfo {pages} {044902} (\bibinfo {year} {2016})},\ \Eprint
  {http://arxiv.org/abs/1506.07164} {arXiv:1506.07164 [hep-ph]} \BibitemShut
  {NoStop}%
\bibitem [{\citenamefont {McNelis}\ \emph {et~al.}(2021)\citenamefont
  {McNelis}, \citenamefont {Bazow},\ and\ \citenamefont
  {Heinz}}]{McNelis:2021zji}%
  \BibitemOpen
  \bibfield  {author} {\bibinfo {author} {\bibfnamefont {M.}~\bibnamefont
  {McNelis}}, \bibinfo {author} {\bibfnamefont {D.}~\bibnamefont {Bazow}}, \
  and\ \bibinfo {author} {\bibfnamefont {U.}~\bibnamefont {Heinz}},\ }\href
  {\doibase 10.1016/j.cpc.2021.108077} {\bibfield  {journal} {\bibinfo
  {journal} {Comput. Phys. Commun.}\ }\textbf {\bibinfo {volume} {267}},\
  \bibinfo {pages} {108077} (\bibinfo {year} {2021})},\ \Eprint
  {http://arxiv.org/abs/2101.02827} {arXiv:2101.02827 [nucl-th]} \BibitemShut
  {NoStop}%
\bibitem [{\citenamefont {Liyanage}\ \emph {et~al.}(2023)\citenamefont
  {Liyanage}, \citenamefont {S\"urer}, \citenamefont {Plumlee}, \citenamefont
  {Wild},\ and\ \citenamefont {Heinz}}]{Liyanage:2023nds}%
  \BibitemOpen
  \bibfield  {author} {\bibinfo {author} {\bibfnamefont {D.}~\bibnamefont
  {Liyanage}}, \bibinfo {author} {\bibfnamefont {O.}~\bibnamefont {S\"urer}},
  \bibinfo {author} {\bibfnamefont {M.}~\bibnamefont {Plumlee}}, \bibinfo
  {author} {\bibfnamefont {S.~M.}\ \bibnamefont {Wild}}, \ and\ \bibinfo
  {author} {\bibfnamefont {U.}~\bibnamefont {Heinz}},\ }\href {\doibase
  10.1103/PhysRevC.108.054905} {\bibfield  {journal} {\bibinfo  {journal}
  {Phys. Rev. C}\ }\textbf {\bibinfo {volume} {108}},\ \bibinfo {pages}
  {054905} (\bibinfo {year} {2023})},\ \Eprint
  {http://arxiv.org/abs/2302.14184} {arXiv:2302.14184 [nucl-th]} \BibitemShut
  {NoStop}%
\bibitem [{\citenamefont {Heinz}\ \emph {et~al.}(2024)\citenamefont {Heinz},
  \citenamefont {Liyanage},\ and\ \citenamefont {Gantenberg}}]{Heinz:2023kzr}%
  \BibitemOpen
  \bibfield  {author} {\bibinfo {author} {\bibfnamefont {U.}~\bibnamefont
  {Heinz}}, \bibinfo {author} {\bibfnamefont {D.}~\bibnamefont {Liyanage}}, \
  and\ \bibinfo {author} {\bibfnamefont {C.}~\bibnamefont {Gantenberg}},\
  }\href {\doibase 10.1051/epjconf/202429605001} {\bibfield  {journal}
  {\bibinfo  {journal} {EPJ Web Conf.}\ }\textbf {\bibinfo {volume} {296}},\
  \bibinfo {pages} {05001} (\bibinfo {year} {2024})},\ \Eprint
  {http://arxiv.org/abs/2311.03306} {arXiv:2311.03306 [nucl-th]} \BibitemShut
  {NoStop}%
\bibitem [{\citenamefont {Zhao}\ \emph {et~al.}(2025)\citenamefont {Zhao},
  \citenamefont {Peng}, \citenamefont {Heinz},\ and\ \citenamefont
  {Song}}]{Zhao:2025jwf}%
  \BibitemOpen
  \bibfield  {author} {\bibinfo {author} {\bibfnamefont {S.}~\bibnamefont
  {Zhao}}, \bibinfo {author} {\bibfnamefont {Y.}~\bibnamefont {Peng}}, \bibinfo
  {author} {\bibfnamefont {U.~W.}\ \bibnamefont {Heinz}}, \ and\ \bibinfo
  {author} {\bibfnamefont {H.}~\bibnamefont {Song}},\ }\href@noop {} {\
  (\bibinfo {year} {2025})},\ \Eprint {http://arxiv.org/abs/2509.03841}
  {arXiv:2509.03841 [nucl-th]} \BibitemShut {NoStop}%
\bibitem [{\citenamefont {Strickland}\ \emph {et~al.}(2016)\citenamefont
  {Strickland}, \citenamefont {Nopoush},\ and\ \citenamefont
  {Ryblewski}}]{Strickland:2015utc}%
  \BibitemOpen
  \bibfield  {author} {\bibinfo {author} {\bibfnamefont {M.}~\bibnamefont
  {Strickland}}, \bibinfo {author} {\bibfnamefont {M.}~\bibnamefont {Nopoush}},
  \ and\ \bibinfo {author} {\bibfnamefont {R.}~\bibnamefont {Ryblewski}},\
  }\href {\doibase 10.1016/j.nuclphysa.2016.02.014} {\bibfield  {journal}
  {\bibinfo  {journal} {Nucl. Phys. A}\ }\textbf {\bibinfo {volume} {956}},\
  \bibinfo {pages} {268} (\bibinfo {year} {2016})},\ \Eprint
  {http://arxiv.org/abs/1512.07334} {arXiv:1512.07334 [nucl-th]} \BibitemShut
  {NoStop}%
\bibitem [{\citenamefont {Chen}\ and\ \citenamefont
  {Shi}(2025)}]{Chen:2024grb}%
  \BibitemOpen
  \bibfield  {author} {\bibinfo {author} {\bibfnamefont {S.}~\bibnamefont
  {Chen}}\ and\ \bibinfo {author} {\bibfnamefont {S.}~\bibnamefont {Shi}},\
  }\href {\doibase 10.1103/PhysRevD.111.014001} {\bibfield  {journal} {\bibinfo
   {journal} {Phys. Rev. D}\ }\textbf {\bibinfo {volume} {111}},\ \bibinfo
  {pages} {014001} (\bibinfo {year} {2025})},\ \Eprint
  {http://arxiv.org/abs/2409.19897} {arXiv:2409.19897 [nucl-th]} \BibitemShut
  {NoStop}%
\bibitem [{\citenamefont {Anderson}\ and\ \citenamefont
  {Witting}(1974)}]{Anderson:1974nyl}%
  \BibitemOpen
  \bibfield  {author} {\bibinfo {author} {\bibfnamefont {J.~L.}\ \bibnamefont
  {Anderson}}\ and\ \bibinfo {author} {\bibfnamefont {H.~R.}\ \bibnamefont
  {Witting}},\ }\href {\doibase 10.1016/0031-8914(74)90355-3} {\bibfield
  {journal} {\bibinfo  {journal} {Physica}\ }\textbf {\bibinfo {volume} {74}},\
  \bibinfo {pages} {466} (\bibinfo {year} {1974})}\BibitemShut {NoStop}%
\bibitem [{\citenamefont {Florkowski}\ \emph
  {et~al.}(2013{\natexlab{b}})\citenamefont {Florkowski}, \citenamefont
  {Ryblewski},\ and\ \citenamefont {Strickland}}]{Florkowski:2013lya}%
  \BibitemOpen
  \bibfield  {author} {\bibinfo {author} {\bibfnamefont {W.}~\bibnamefont
  {Florkowski}}, \bibinfo {author} {\bibfnamefont {R.}~\bibnamefont
  {Ryblewski}}, \ and\ \bibinfo {author} {\bibfnamefont {M.}~\bibnamefont
  {Strickland}},\ }\href {\doibase 10.1103/PhysRevC.88.024903} {\bibfield
  {journal} {\bibinfo  {journal} {Phys. Rev. C}\ }\textbf {\bibinfo {volume}
  {88}},\ \bibinfo {pages} {024903} (\bibinfo {year} {2013}{\natexlab{b}})},\
  \Eprint {http://arxiv.org/abs/1305.7234} {arXiv:1305.7234 [nucl-th]}
  \BibitemShut {NoStop}%
\bibitem [{\citenamefont {Bazavov}\ \emph {et~al.}(2014)\citenamefont {Bazavov}
  \emph {et~al.}}]{HotQCD:2014kol}%
  \BibitemOpen
  \bibfield  {author} {\bibinfo {author} {\bibfnamefont {A.}~\bibnamefont
  {Bazavov}} \emph {et~al.} (\bibinfo {collaboration} {HotQCD}),\ }\href
  {\doibase 10.1103/PhysRevD.90.094503} {\bibfield  {journal} {\bibinfo
  {journal} {Phys. Rev. D}\ }\textbf {\bibinfo {volume} {90}},\ \bibinfo
  {pages} {094503} (\bibinfo {year} {2014})},\ \Eprint
  {http://arxiv.org/abs/1407.6387} {arXiv:1407.6387 [hep-lat]} \BibitemShut
  {NoStop}%
\bibitem [{\citenamefont {Borsanyi}\ \emph {et~al.}(2016)\citenamefont
  {Borsanyi} \emph {et~al.}}]{Borsanyi:2016ksw}%
  \BibitemOpen
  \bibfield  {author} {\bibinfo {author} {\bibfnamefont {S.}~\bibnamefont
  {Borsanyi}} \emph {et~al.},\ }\href {\doibase 10.1038/nature20115} {\bibfield
   {journal} {\bibinfo  {journal} {Nature}\ }\textbf {\bibinfo {volume}
  {539}},\ \bibinfo {pages} {69} (\bibinfo {year} {2016})},\ \Eprint
  {http://arxiv.org/abs/1606.07494} {arXiv:1606.07494 [hep-lat]} \BibitemShut
  {NoStop}%
\bibitem [{\citenamefont {Grad}(1949)}]{Grad:1949zza}%
  \BibitemOpen
  \bibfield  {author} {\bibinfo {author} {\bibfnamefont {H.}~\bibnamefont
  {Grad}},\ }\href {\doibase 10.1002/cpa.3160020403} {\bibfield  {journal}
  {\bibinfo  {journal} {Commun. Pure Appl. Math.}\ }\textbf {\bibinfo {volume}
  {2}},\ \bibinfo {pages} {331} (\bibinfo {year} {1949})}\BibitemShut {NoStop}%
\bibitem [{\citenamefont {Baier}\ \emph {et~al.}(2006)\citenamefont {Baier},
  \citenamefont {Romatschke},\ and\ \citenamefont {Wiedemann}}]{Baier:2006um}%
  \BibitemOpen
  \bibfield  {author} {\bibinfo {author} {\bibfnamefont {R.}~\bibnamefont
  {Baier}}, \bibinfo {author} {\bibfnamefont {P.}~\bibnamefont {Romatschke}}, \
  and\ \bibinfo {author} {\bibfnamefont {U.~A.}\ \bibnamefont {Wiedemann}},\
  }\href {\doibase 10.1103/PhysRevC.73.064903} {\bibfield  {journal} {\bibinfo
  {journal} {Phys. Rev. C}\ }\textbf {\bibinfo {volume} {73}},\ \bibinfo
  {pages} {064903} (\bibinfo {year} {2006})},\ \Eprint
  {http://arxiv.org/abs/hep-ph/0602249} {arXiv:hep-ph/0602249} \BibitemShut
  {NoStop}%
\bibitem [{\citenamefont {Baier}\ \emph {et~al.}(2008)\citenamefont {Baier},
  \citenamefont {Romatschke}, \citenamefont {Son}, \citenamefont {Starinets},\
  and\ \citenamefont {Stephanov}}]{Baier:2007ix}%
  \BibitemOpen
  \bibfield  {author} {\bibinfo {author} {\bibfnamefont {R.}~\bibnamefont
  {Baier}}, \bibinfo {author} {\bibfnamefont {P.}~\bibnamefont {Romatschke}},
  \bibinfo {author} {\bibfnamefont {D.~T.}\ \bibnamefont {Son}}, \bibinfo
  {author} {\bibfnamefont {A.~O.}\ \bibnamefont {Starinets}}, \ and\ \bibinfo
  {author} {\bibfnamefont {M.~A.}\ \bibnamefont {Stephanov}},\ }\href {\doibase
  10.1088/1126-6708/2008/04/100} {\bibfield  {journal} {\bibinfo  {journal}
  {JHEP}\ }\textbf {\bibinfo {volume} {04}},\ \bibinfo {pages} {100} (\bibinfo
  {year} {2008})},\ \Eprint {http://arxiv.org/abs/0712.2451} {arXiv:0712.2451
  [hep-th]} \BibitemShut {NoStop}%
\bibitem [{\citenamefont {Betz}\ \emph {et~al.}(2009)\citenamefont {Betz},
  \citenamefont {Henkel},\ and\ \citenamefont {Rischke}}]{Betz:2008me}%
  \BibitemOpen
  \bibfield  {author} {\bibinfo {author} {\bibfnamefont {B.}~\bibnamefont
  {Betz}}, \bibinfo {author} {\bibfnamefont {D.}~\bibnamefont {Henkel}}, \ and\
  \bibinfo {author} {\bibfnamefont {D.~H.}\ \bibnamefont {Rischke}},\ }\href
  {\doibase 10.1016/j.ppnp.2008.12.018} {\bibfield  {journal} {\bibinfo
  {journal} {Prog. Part. Nucl. Phys.}\ }\textbf {\bibinfo {volume} {62}},\
  \bibinfo {pages} {556} (\bibinfo {year} {2009})},\ \Eprint
  {http://arxiv.org/abs/0812.1440} {arXiv:0812.1440 [nucl-th]} \BibitemShut
  {NoStop}%
\bibitem [{\citenamefont {Denicol}\ \emph
  {et~al.}(2012{\natexlab{a}})\citenamefont {Denicol}, \citenamefont {Niemi},
  \citenamefont {Molnar},\ and\ \citenamefont {Rischke}}]{Denicol:2012cn}%
  \BibitemOpen
  \bibfield  {author} {\bibinfo {author} {\bibfnamefont {G.~S.}\ \bibnamefont
  {Denicol}}, \bibinfo {author} {\bibfnamefont {H.}~\bibnamefont {Niemi}},
  \bibinfo {author} {\bibfnamefont {E.}~\bibnamefont {Molnar}}, \ and\ \bibinfo
  {author} {\bibfnamefont {D.~H.}\ \bibnamefont {Rischke}},\ }\href {\doibase
  10.1103/PhysRevD.85.114047} {\bibfield  {journal} {\bibinfo  {journal} {Phys.
  Rev. D}\ }\textbf {\bibinfo {volume} {85}},\ \bibinfo {pages} {114047}
  (\bibinfo {year} {2012}{\natexlab{a}})},\ \bibinfo {note} {[Erratum:
  Phys.Rev.D 91, 039902 (2015)]},\ \Eprint {http://arxiv.org/abs/1202.4551}
  {arXiv:1202.4551 [nucl-th]} \BibitemShut {NoStop}%
\bibitem [{\citenamefont {Denicol}\ \emph
  {et~al.}(2012{\natexlab{b}})\citenamefont {Denicol}, \citenamefont
  {Moln{\'a}r}, \citenamefont {Niemi},\ and\ \citenamefont
  {Rischke}}]{Denicol:2012es}%
  \BibitemOpen
  \bibfield  {author} {\bibinfo {author} {\bibfnamefont {G.~S.}\ \bibnamefont
  {Denicol}}, \bibinfo {author} {\bibfnamefont {E.}~\bibnamefont {Moln{\'a}r}},
  \bibinfo {author} {\bibfnamefont {H.}~\bibnamefont {Niemi}}, \ and\ \bibinfo
  {author} {\bibfnamefont {D.~H.}\ \bibnamefont {Rischke}},\ }\href {\doibase
  10.1140/epja/i2012-12170-x} {\bibfield  {journal} {\bibinfo  {journal} {Eur.
  Phys. J. A}\ }\textbf {\bibinfo {volume} {48}},\ \bibinfo {pages} {170}
  (\bibinfo {year} {2012}{\natexlab{b}})},\ \Eprint
  {http://arxiv.org/abs/1206.1554} {arXiv:1206.1554 [nucl-th]} \BibitemShut
  {NoStop}%
\bibitem [{\citenamefont {Israel}\ and\ \citenamefont
  {Stewart}(1979)}]{Israel:1979wp}%
  \BibitemOpen
  \bibfield  {author} {\bibinfo {author} {\bibfnamefont {W.}~\bibnamefont
  {Israel}}\ and\ \bibinfo {author} {\bibfnamefont {J.~M.}\ \bibnamefont
  {Stewart}},\ }\href {\doibase 10.1016/0003-4916(79)90130-1} {\bibfield
  {journal} {\bibinfo  {journal} {Annals Phys.}\ }\textbf {\bibinfo {volume}
  {118}},\ \bibinfo {pages} {341} (\bibinfo {year} {1979})}\BibitemShut
  {NoStop}%
\bibitem [{\citenamefont {Muronga}\ and\ \citenamefont
  {Rischke}(2004)}]{Muronga:2004sf}%
  \BibitemOpen
  \bibfield  {author} {\bibinfo {author} {\bibfnamefont {A.}~\bibnamefont
  {Muronga}}\ and\ \bibinfo {author} {\bibfnamefont {D.~H.}\ \bibnamefont
  {Rischke}},\ }\href@noop {} {\  (\bibinfo {year} {2004})},\ \Eprint
  {http://arxiv.org/abs/nucl-th/0407114} {arXiv:nucl-th/0407114} \BibitemShut
  {NoStop}%
\bibitem [{\citenamefont {Jaiswal}(2013)}]{Jaiswal:2013npa}%
  \BibitemOpen
  \bibfield  {author} {\bibinfo {author} {\bibfnamefont {A.}~\bibnamefont
  {Jaiswal}},\ }\href {\doibase 10.1103/PhysRevC.87.051901} {\bibfield
  {journal} {\bibinfo  {journal} {Phys. Rev. C}\ }\textbf {\bibinfo {volume}
  {87}},\ \bibinfo {pages} {051901} (\bibinfo {year} {2013})},\ \Eprint
  {http://arxiv.org/abs/1302.6311} {arXiv:1302.6311 [nucl-th]} \BibitemShut
  {NoStop}%
\bibitem [{\citenamefont {Moln\'ar}\ \emph {et~al.}(2014)\citenamefont
  {Moln\'ar}, \citenamefont {Niemi}, \citenamefont {Denicol},\ and\
  \citenamefont {Rischke}}]{Molnar:2013lta}%
  \BibitemOpen
  \bibfield  {author} {\bibinfo {author} {\bibfnamefont {E.}~\bibnamefont
  {Moln\'ar}}, \bibinfo {author} {\bibfnamefont {H.}~\bibnamefont {Niemi}},
  \bibinfo {author} {\bibfnamefont {G.~S.}\ \bibnamefont {Denicol}}, \ and\
  \bibinfo {author} {\bibfnamefont {D.~H.}\ \bibnamefont {Rischke}},\ }\href
  {\doibase 10.1103/PhysRevD.89.074010} {\bibfield  {journal} {\bibinfo
  {journal} {Phys. Rev. D}\ }\textbf {\bibinfo {volume} {89}},\ \bibinfo
  {pages} {074010} (\bibinfo {year} {2014})},\ \Eprint
  {http://arxiv.org/abs/1308.0785} {arXiv:1308.0785 [nucl-th]} \BibitemShut
  {NoStop}%
\bibitem [{\citenamefont {Ambrus}\ \emph
  {et~al.}(2022{\natexlab{a}})\citenamefont {Ambrus}, \citenamefont
  {Moln{\'a}r},\ and\ \citenamefont {Rischke}}]{Ambrus:2022vif}%
  \BibitemOpen
  \bibfield  {author} {\bibinfo {author} {\bibfnamefont {V.~E.}\ \bibnamefont
  {Ambrus}}, \bibinfo {author} {\bibfnamefont {E.}~\bibnamefont {Moln{\'a}r}},
  \ and\ \bibinfo {author} {\bibfnamefont {D.~H.}\ \bibnamefont {Rischke}},\
  }\href {\doibase 10.1103/PhysRevD.106.076005} {\bibfield  {journal} {\bibinfo
   {journal} {Phys. Rev. D}\ }\textbf {\bibinfo {volume} {106}},\ \bibinfo
  {pages} {076005} (\bibinfo {year} {2022}{\natexlab{a}})},\ \Eprint
  {http://arxiv.org/abs/2207.05670} {arXiv:2207.05670 [nucl-th]} \BibitemShut
  {NoStop}%
\bibitem [{\citenamefont {Karpenko}\ \emph {et~al.}(2014)\citenamefont
  {Karpenko}, \citenamefont {Huovinen},\ and\ \citenamefont
  {Bleicher}}]{Karpenko:2013wva}%
  \BibitemOpen
  \bibfield  {author} {\bibinfo {author} {\bibfnamefont {I.}~\bibnamefont
  {Karpenko}}, \bibinfo {author} {\bibfnamefont {P.}~\bibnamefont {Huovinen}},
  \ and\ \bibinfo {author} {\bibfnamefont {M.}~\bibnamefont {Bleicher}},\
  }\href {\doibase 10.1016/j.cpc.2014.07.010} {\bibfield  {journal} {\bibinfo
  {journal} {Comput. Phys. Commun.}\ }\textbf {\bibinfo {volume} {185}},\
  \bibinfo {pages} {3016} (\bibinfo {year} {2014})},\ \Eprint
  {http://arxiv.org/abs/1312.4160} {arXiv:1312.4160 [nucl-th]} \BibitemShut
  {NoStop}%
\bibitem [{\citenamefont {Borghini}\ \emph {et~al.}(2023)\citenamefont
  {Borghini}, \citenamefont {Borrell}, \citenamefont {Feld}, \citenamefont
  {Roch}, \citenamefont {Schlichting},\ and\ \citenamefont
  {Werthmann}}]{Borghini:2022iym}%
  \BibitemOpen
  \bibfield  {author} {\bibinfo {author} {\bibfnamefont {N.}~\bibnamefont
  {Borghini}}, \bibinfo {author} {\bibfnamefont {M.}~\bibnamefont {Borrell}},
  \bibinfo {author} {\bibfnamefont {N.}~\bibnamefont {Feld}}, \bibinfo {author}
  {\bibfnamefont {H.}~\bibnamefont {Roch}}, \bibinfo {author} {\bibfnamefont
  {S.}~\bibnamefont {Schlichting}}, \ and\ \bibinfo {author} {\bibfnamefont
  {C.}~\bibnamefont {Werthmann}},\ }\href {\doibase
  10.1103/PhysRevC.107.034905} {\bibfield  {journal} {\bibinfo  {journal}
  {Phys. Rev. C}\ }\textbf {\bibinfo {volume} {107}},\ \bibinfo {pages}
  {034905} (\bibinfo {year} {2023})},\ \Eprint
  {http://arxiv.org/abs/2209.01176} {arXiv:2209.01176 [hep-ph]} \BibitemShut
  {NoStop}%
\bibitem [{\citenamefont {Ambrus}\ \emph
  {et~al.}(2022{\natexlab{b}})\citenamefont {Ambrus}, \citenamefont
  {Schlichting},\ and\ \citenamefont {Werthmann}}]{Ambrus:2021fej}%
  \BibitemOpen
  \bibfield  {author} {\bibinfo {author} {\bibfnamefont {V.~E.}\ \bibnamefont
  {Ambrus}}, \bibinfo {author} {\bibfnamefont {S.}~\bibnamefont {Schlichting}},
  \ and\ \bibinfo {author} {\bibfnamefont {C.}~\bibnamefont {Werthmann}},\
  }\href {\doibase 10.1103/PhysRevD.105.014031} {\bibfield  {journal} {\bibinfo
   {journal} {Phys. Rev. D}\ }\textbf {\bibinfo {volume} {105}},\ \bibinfo
  {pages} {014031} (\bibinfo {year} {2022}{\natexlab{b}})},\ \Eprint
  {http://arxiv.org/abs/2109.03290} {arXiv:2109.03290 [hep-ph]} \BibitemShut
  {NoStop}%
\bibitem [{\citenamefont {Ambrus}\ \emph
  {et~al.}(2023{\natexlab{c}})\citenamefont {Ambrus}, \citenamefont
  {Schlichting},\ and\ \citenamefont {Werthmann}}]{Ambrus:2022oji}%
  \BibitemOpen
  \bibfield  {author} {\bibinfo {author} {\bibfnamefont {V.~E.}\ \bibnamefont
  {Ambrus}}, \bibinfo {author} {\bibfnamefont {S.}~\bibnamefont {Schlichting}},
  \ and\ \bibinfo {author} {\bibfnamefont {C.}~\bibnamefont {Werthmann}},\
  }\href {\doibase 10.5506/APhysPolBSupp.16.1-A32} {\bibfield  {journal}
  {\bibinfo  {journal} {Acta Phys. Polon. Supp.}\ }\textbf {\bibinfo {volume}
  {16}},\ \bibinfo {pages} {1} (\bibinfo {year} {2023}{\natexlab{c}})},\
  \Eprint {http://arxiv.org/abs/2207.12789} {arXiv:2207.12789 [hep-ph]}
  \BibitemShut {NoStop}%
\bibitem [{\citenamefont {Vredevoogd}\ and\ \citenamefont
  {Pratt}(2009)}]{Vredevoogd:2008id}%
  \BibitemOpen
  \bibfield  {author} {\bibinfo {author} {\bibfnamefont {J.}~\bibnamefont
  {Vredevoogd}}\ and\ \bibinfo {author} {\bibfnamefont {S.}~\bibnamefont
  {Pratt}},\ }\href {\doibase 10.1103/PhysRevC.79.044915} {\bibfield  {journal}
  {\bibinfo  {journal} {Phys. Rev. C}\ }\textbf {\bibinfo {volume} {79}},\
  \bibinfo {pages} {044915} (\bibinfo {year} {2009})},\ \Eprint
  {http://arxiv.org/abs/0810.4325} {arXiv:0810.4325 [nucl-th]} \BibitemShut
  {NoStop}%
\bibitem [{\citenamefont {Peng}(2025)}]{peng_2025_Zenodo}%
  \BibitemOpen
  \bibfield  {author} {\bibinfo {author} {\bibfnamefont {Y.}~\bibnamefont
  {Peng}},\ }\href@noop {} {\enquote {\bibinfo {title} {Plot data for {\it
  extended applicability domain of viscous anisotropic hydrodynamics in (2+1)-d
  bjorken flow with transverse expansion}},}\ } (\bibinfo {year} {2025}),\
  \bibinfo {note} {{DOI:}
  \href{https://doi.org/10.5281/zenodo.17207917}{10.5281/zenodo.17207917}}\BibitemShut
  {NoStop}%
\bibitem [{\citenamefont {Giacalone}\ \emph {et~al.}(2019)\citenamefont
  {Giacalone}, \citenamefont {Mazeliauskas},\ and\ \citenamefont
  {Schlichting}}]{Giacalone:2019ldn}%
  \BibitemOpen
  \bibfield  {author} {\bibinfo {author} {\bibfnamefont {G.}~\bibnamefont
  {Giacalone}}, \bibinfo {author} {\bibfnamefont {A.}~\bibnamefont
  {Mazeliauskas}}, \ and\ \bibinfo {author} {\bibfnamefont {S.}~\bibnamefont
  {Schlichting}},\ }\href {\doibase 10.1103/PhysRevLett.123.262301} {\bibfield
  {journal} {\bibinfo  {journal} {Phys. Rev. Lett.}\ }\textbf {\bibinfo
  {volume} {123}},\ \bibinfo {pages} {262301} (\bibinfo {year} {2019})},\
  \Eprint {http://arxiv.org/abs/1908.02866} {arXiv:1908.02866 [hep-ph]}
  \BibitemShut {NoStop}%
\end{thebibliography}%

\end{document}